\documentclass{article}
\usepackage{amsmath,amsthm,amssymb,amscd,epsfig,indentfirst,latexsym}
\input amssym.def
\input amssym.tex

\numberwithin{equation}{section}

\newtheorem{theoreme}{Theorem }[section]
\newtheorem{proposition}[theoreme]{Proposition}
\newtheorem{lemma}[theoreme]{Lemma}
\newtheorem{definition}[theoreme]{Definition}
\newtheorem{corollary}[theoreme]{Corollary}
\newtheorem{remark}[theoreme]{Remark}
\newtheorem{example}[theoreme]{Example}
\newtheorem{assumption}{Assumption}[section]

\newcommand{\beq}{\begin{equation}}
\newcommand{\eeq}{\end{equation}}
\newcommand{\ben}{\begin{arabicenumerate}}
\newcommand{\een}{\end{arabicenumerate}}
\newcommand{\bex}{\begin{example}}
\newcommand{\eex}{\end{example}}
\newcommand{\ber}{\begin{remark}}
\newcommand{\eer}{\end{remark}}
\def\bel{\begin{lemma}}
\def\eel{\end{lemma}}
\def\bet{\begin{theoreme}}
\def\eet{\end{theoreme}}
\def\bed{\begin{definition}}
\def\eed{\end{definition}}
\def\bep{\begin{proposition}}
\def\eep{\end{proposition}}
\def\bec{\begin{corollary}}
\def\eec{\end{corollary}}
\def\bea{\begin{assumption}}
\def\eea{\end{assumption}}

\newcommand{\comment}[1]{}

\newcounter{smallarabics}
\newenvironment{arabicenumerate}
{\begin{list}{{\normalfont\textrm{(\arabic{smallarabics})}}}
  {\usecounter{smallarabics}\setlength{\itemindent}{0cm}
   \setlength{\leftmargin}{5ex}\setlength{\labelwidth}{4ex}
   \setlength{\topsep}{0.75\parsep}\setlength{\partopsep}{0ex}
   \setlength{\itemsep}{0ex}}}
{\end{list}}

\def\rr{{\mathbb R}}

\def\cc{{\mathbb C}}
\def\nn{{\mathbb N}}

 \def\cB{{\cal B}} 
\def\cD{{\cal D}}  
 \def\cH{{\cal H}} 
  
 \def\cN{{\cal N}}

\def\cV{{\cal V}}  
\def\cY{{\cal Y}}

\def\fh{{\frak h}}

\def\d{{\rm d}}
\def\e{{\rm e}}
\def\i{{\rm i}}

\def\s{{\rm s}}

\def\FDom{{\rm Dom}}

\def\(form){{\rm (form)}}

\def\Im{{\rm Im}}
\def\Re{{\rm Re}}

\def\fin{{\rm fin}}

\def\Tr{{\rm Tr}}
\def\det{{\rm det}}
\def\Ker{{\rm Ker}}

\def\Op{{\rm Op}}
\def\ren{{\rm ren}}
\def\nat{{\rm nat}}

\def\proof{\noindent{\bf Proof.}\ \ }
\def\qed{$\Box$\medskip}

\def\slim{{\rm s-}\lim}

\begin{document}

\title{Bogoliubov Hamiltonians and one-parameter groups of Bogoliubov transformations}

\author{Laurent Bruneau\footnote{CPT-CNRS, UMR 6207, Universit\'e du Sud, Toulon-Var, B.P. 20132,
F-83957 La Garde Cedex, France. Email: bruneau@cpt.univ-mrs.fr}\, and Jan Derezi\'nski\footnote{Department of Mathematical Methods in Physics,
Warsaw University, Ho\.{z}a 74, 00-682, Warszawa, Poland. Email: Jan.Derezinski@fuw.edu.pl}}

\maketitle

\begin{abstract} 
On the bosonic Fock space, a family of Bogoliubov transformations
corresponding to a strongly continuous one-parameter group of
symplectic maps $(R(t))_{t\in\rr}$ is considered. Under suitable
assumptions on the generator $A$ of this group, which guarantee that
the induced representations of CCR are unitarily equivalent for all
time $t,$ it is known that the unitary operator $U_\nat(t)$ which
implement this transformation gives a projective unitary
representation of $R(t)$. Under rather general assumptions on the generator
$A,$ we prove that the corresponding Bogoliubov transformations can be
implemented by a one-parameter group $U(t)$ of unitary operators. The
generator of $U(t)$ will be called a Bogoliubov Hamiltonian. We will
introduce two kinds of Bogoliubov Hamiltonians (type I and II) and
give conditions so that they are well defined.

\end{abstract}

\noindent{\small {\bf Acknowledgments.} 
The research of both authors
 was supported by  the Postdoctoral Training Program 
HPRN-CT-2002-0277. The first author would like to thank Warsaw University, 
the CRM of Montreal and Institut Fourier in Grenoble where part of
this work has been done. The research of the second author
 was also supported by the Polish grants
 SPUB127 and  2 P03A 027 25 and was partly done during his visit to the Erwin
 Schr\"odinger Institute as a Senior Research Fellow.
}


\section{Introduction}\label{sec:intro}

Given a real symplectic space $\cY$ (with a symplectic form $\sigma$) and a complex Hilbert space $\cH,$ a representation of the Canonical Commutation Relations (CCR) over $\cY$ in $\cH$ is a map $\cY\ni y \mapsto W(y)\in U(\cH)$ (the unitary operators on $\cH$) such that
\beq\label{def:reprccr2}
W(y)W(y')=\e^{-\frac{\i}{2}\sigma(y,y')}W(y+y'), \quad \forall y,y'\in\cY.
\eeq
These representations arise naturally in the study of bosonic systems (\emph{e.g.} a free Bose field).

For any symplectic map $R$ on $\cY$ (\emph{i.e.}
$\sigma(Ry,Ry')=\sigma(y,y')$ for all $y,y'$), the map  
\[
\cY \ni y \mapsto W_R(y):=W(Ry)
\]
is also a representation of CCR. The tranformation of $W(y)$ into $W_R(y)$ 
is often  called
 the Bogoliubov transformation. The question is whether these two
representations ($W$ and $W_R$) are unitarily equivalent, i.e. is there a unitary operator $U$ on $\cH$ such that for all $y$ in $\cY,$ $UW(y)U^{-1}=W_R(y).$

In our paper we consider the so-called
Fock representation, \emph{i.e.} the Hilbert space $\cH$ is a bosonic
Fock space $\Gamma_\s(\fh)$ and the operators $W(y)$ are the usual  Weyl
operators. In this case the symplectic spaces is 
$\cY:= \{ (f, \bar{f}) , f \in\fh\},$
where $f\mapsto \bar{f}$ is a conjugation (i.e. 
an antilinear involution) on the Hilbert space $\fh.$ The symplectic form on $\cY$ will be the imaginary part of the scalar product on $\fh$ (\emph{i.e.} $\sigma((f,\bar{f}),(g,\bar{g}))=\Im \langle f|g\rangle$).

Consider a real symplectic map $R$ on $\cY.$ One can write it as a $2\times2$ matrix,
$
R= \left(\begin{array}{cc} P & \bar{Q} \\ Q & \bar{P} \end{array}\right),
$
where $P$ and $Q$ are bounded operators on $\fh$ which satisfy some conditions (see (\ref{eq:symplcond})). We also introduce
$
J=\left(\begin{array}{cc} \i & 0 \\ 0 & -\i \end{array}\right).
$
Finally, let $W(f)$ denote the Weyl operators on the bosonic Fock
space $\Gamma_\s(\fh).$ 
In that setting, the question of unitarily equivalence has been solved by Shale \cite{Sh}: the representations of $W$ and $W_R$ are unitarily equivalent if and only if the operator $[R,J]$
is a Hilbert-Schmidt operator on $\fh\oplus\fh$, which is equivalent
to say that the operator $Q$ is Hilbert-Schmidt on $\fh.$ Moreover, if
this condition holds, then the operator, which we will call
 the natural Bogoliubov implementer, 
\[
U_\nat:= {\rm det} (1-K^*K)^{1/4} \e^{-\frac{1}{2}a^*(K)}\Gamma((P^{-1})^*) \e^{-\frac{1}{2}a(L)},
\]
with $K=\overline{QP^{-1}}$ and $L=-P^{-1}\bar{Q},$ extends to a unitary
operator on $\Gamma_\s(\fh)$ and satisfies
\cite{IH,Ru1,Ru2}: $\forall f\in\fh, \, W_R(f)=U_\nat W(f)U_\nat^*$. Here, $a(L)$ and $a^*(K)$ denote the ``quadratic'' annihilation and creation operators associated to the Hilbert-Schmidt operators $K$ and $L$ via the natural identification between Hilbert-Schmidt operators on $\fh$ and vectors in $\fh\otimes\fh$ (Section \ref{ssec:doubleaa*}).

Suppose $(R(t))_{t\in \rr}$ is a group of symplectic maps such that,
for all $t,$ $[R(t),J]$ is Hilbert-Schmidt. We can then define the
operators $U_\nat(t)$ for all $t$. In general
 $(U_\nat(t))_{t\in\rr}$ will not be a one-parameter group as well.
 Since $(R(t))_{t\in \rr}$ is a one-parameter group and the Weyl representation $W$ is irreducible we know that
\[
U_\nat(t)U_\nat(s)=\e^{\i\rho(t,s)}U_\nat(t+s).
\]
Clearly, for any $\theta(t)\in\rr,$ $U(t):=\e^{\i\theta(t)}U_\nat(t)$ also
intertwines $W$ and $W_{R(t)}$.
A suitable choice of the phase
$\theta(t)$ may give rise to a strongly continuous 
unitary group
$U(t)$. When such a unitary group exists, $R(t)$ will be called
unitarily implementable and its selfadjoint generator $H$
 a Bogoliubov Hamiltonian. Note that
the irreducibility of $W$ guarantees that all the Bogoliubov Hamiltonians associated
to a given symplectic group are equal up to a constant. 

There are at least 
two natural choices for this constant, corresponding to two
distinguished classes of Bogoliubov Hamiltonians, which we call type I
and type II. The type I Bogoliubov Hamiltonian is such that its
expectation value on the Fock vacuum vanishes. The type II Bogoliubov
Hamiltonian  is such that its infimum is zero. We will see that
 such choices are not
always possible, i.e. one of these distinguished Bogoliubov
Hamiltonians (or both) may not exist, even if $R(t)$ is unitarily
implementable. 

Let 
$
A=\i\left(\begin{array}{cc} h & -v \\ \bar{v} & -\bar{h}
  \end{array}\right)
$
denote the generator of the symplectic group $R(t)$. 
$R(t)$ is symplectic for all $t$ if and only if $h$ is selfadjoint and $v^*=\bar{v}.$
We will see that, at least formally, the Bogoliubov Hamiltonians
associated to $R(t)$ are given by 
\beq\label{intro:ham}
H:=\d\Gamma(h)+\frac{1}{2}(a^*(v)+a(v))+c.
\eeq
Here $c$ is a constant, which may be infinite -- this means that one may have
 to
perform an approrpriate renormalization (see Section \ref{sec:example} for a
concrete example).

 It is easy to see that the constant $c_I$
corresponding to the type I Bogoliubov Hamiltonian is zero. The
constant $c_{II}$ (at least in the case where
$\fh$ is finite dimensional) is given by 
$$
c_{II}=-\frac{1}{4}\Tr \left[ \left( \begin{array}{cc} \bar{h}^2-\bar{v}v & \bar{h}\bar{v}-\bar{v}h
 \\ hv-v\bar{h}  & h^2-v\bar{v}  \end{array}\right)^{1/2}-\left( \begin{array}{cc} \bar{h} & 0 \\ 0
 & h \end{array}\right) \right].
$$ 

The unitary group generated by the type I Hamiltonian can be written explicitly:\beq\label{intro:group}
U_I(t):=\e^{\i tH_I}={\rm det}(\bar{P}(t)\e^{\i t\bar{h}})^{-1/2}\e^{-\frac{1}{2}a^*(K(t))}\Gamma((P^{-1}(t))^*) \e^{-\frac{1}{2}a(L(t))}.
\eeq

In our paper we give conditions on the generator $A$ of $R(t)$
 guaranteeing so that $R(t)$
 is unitarily implementable and conditions which ensure
that Bogoliubov Hamiltonians of type I, resp. type II, are well defined.

We prove that, if for all $t$ the operator
\beq\label{intro:v(t)}
v(t):=\int_0^t \e^{\i\tau h}v\e^{\i\tau \bar{h}}\d\tau
\eeq
is Hilbert-Schmidt such that its Hilbert-Schmidt norm is locally
integrable and continuous at $t=0$, then $R(t)$ is unitarily implementable. 

To guarantee the existence of type I, we need some additional
assumption on $A$, namely: the operator $\bar{v}v(t)$ is trace class,
and its trace norm is locally integrable and continuous at
$t=0$. Under this condition, we prove that the operators $U(t)$ defined by
(\ref{intro:group}) with $c=0$ form a strongly continuous unitary
group. Note that this does note require $v$ to be Hilbert-Schmidt and
allows to give a meaning to the formal operator (\ref{intro:ham}) in a more general situation.
On the other hand, if $v$ is Hilbert-Schmidt, then the above
assumptions are satisfied and hence the Bogoliubov Hamiltonian of type I
exists. Moreover, we then prove that the a priori formal expression
(\ref{intro:ham}) indeed defines an essentially selfadjoint operator.

The selfadjointness of $H$ is not obvious, and to our knowledge there
is no (rigorous) proof of it. We would like to emphasize the fact that
$v$ is naturally associated to an element of $\Gamma^2_\s(\fh)$ and not
$\Gamma^1_\s(\fh)=\fh,$ so that the ``perturbation'' $a^*(v)+a(v)$ has really
to be thought as an operator quadratic, and not linear, in $a$ and $a^*$.

In order to study Bogoliubov Hamiltonians of type II, one needs to
compute the infimum of operators $H$ of the form (\ref{intro:ham}). In
particular, the Bogoliuobov Hamiltonian of type II is well defined if
and only if $R(t)$ has bounded from below Bogoliubov Hamiltonians.

If
$\fh$ has finite dimension, we will prove that $H$ is bounded from
below (and compute its infimum) if and only if, for all $f\in\fh$, 
$$
\langle f|hf\rangle+\langle
\bar{f}|\bar{h}\bar{f}\rangle+\langle f|v\bar{f}\rangle+\langle
\bar{f}|\bar{v} f\rangle \geq 0.
$$

When $h$ is positive, we also give a condition under which  the ``perturbation'' $a^*(v)+a(v)$ is relatively bounded
with respect to $\d\Gamma(h)$, and we give an upper bound on this
relative bound. We thus get another class of
symplectic groups for which both type I and type II Bogoliubov
Hamiltonians exist.

Finally, we study completely the simple (non
trivial) following situation: $\fh=L^2(\nn)$ and the operators $h$ and
$v$ are both diagonal with respect to the canonical basis of
$L^2(\nn),$ i.e. $h=\sum h_n |e_n\rangle\langle e_n|$ and $v=\sum v_n |e_n\rangle\langle e_n|.$
We prove that $R(t)$ is unitarily implementable if and
only if $\sum \frac{|v_n|^2}{1+h_n^2}<+\infty.$ 
Then we prove that the Bogoliubov
Hamiltonian of type I, resp. type II, is well defined if and only if
$\sum \frac{|v_n|^2}{1+|h_n|}<+\infty$, resp. $\sum_{|h_n|\leq 1}
\frac{|v_n|^2}{|h_n|}<+\infty$. In particular one can see 
that all kinds of situation can occur: neither type I nor type II exist,
type I exists but not type II, etc.

We end this introduction with a few comments on the related results
which exist in the litterature.
In \cite{Be}, under the same conditions on $v(t)$ (see
(\ref{intro:v(t)}) and below), it
is proved that the operators $U_I(t)$  are well defined, and that they
form a one-parameter group of unitary operators whose generator is
given by (\ref{intro:ham}) provided the latter makes sense as an
essentially selfadjoint operator. The author also proves this
essential selfadjointness when $v$ is Hilbert-Schmidt. However, the
proofs at some places are not quite complete (the proof
of essential selfadjointness for instance is not completely
rigorous).  
Similar results  are also obtained in \cite{Ne} but the author considers only the
case where $v$ is Hilbert-Schmidt and partially relies on the results
of \cite{Be}, such  
as for the essential selfadjointness of Bogoliubov Hamiltonians.

More recently, in \cite{IH} the authors have also considered the
question of finding a unitary group $\e^{\i tH}$ which intertwines $W$
and $W_{R(t)}$ but only for norm continuous symplectic groups.

Finally, we would like to mention that 
similar ``quadratic operators'' as (\ref{intro:ham}) have been studied
in \emph{e.g.} \cite{A,AY}. However, the authors use the field operator
$\phi(f)=a(f)+a^*(f)$ instead of the annihilation/creation
operators. Namely, if $\Gamma_\s^2(\fh)\ni v=\sum \lambda_n e_n\otimes e_n$ where $(e_n)_n$ is an orthonormal basis of $\fh$ and $\lambda_n$ are positive numbers, then the operator $\sum \lambda_n \phi(e_n)\phi(e_n)$ is considered, while we use operators of the form $\sum \lambda_n a^*(e_n)a^*(e_n).$ In particular, the use of the field operators in the previous sum leads to quadratic expressions which are not normal ordered. Therefore, in order to make them well defined, one has to impose that  $v$ is actually trace class.


\section{Fock spaces and representation of the CCR}\label{sec:fockandccr}


\subsection{Generalities on the Fock space}\label{ssec:generfock}

Let $\fh$ be a Hilbert space. We denote by $\Gamma_\s(\fh)$ the bosonic Fock space over the one-particle space $\fh,$ 
\[
\Gamma_\s(\fh):= \bigoplus_{n=0}^{\infty} \Gamma^n_\s(\fh),
\]
where $\Gamma^n_\s(\fh):=\otimes^n_\s \fh$ denotes the symmetric $n-$fold tensor product of $\fh$ with the convention $\otimes^0_\s \fh=\cc.$ $\Omega:= (1, 0,\cdots )$ will denote the vacuum vector and 
$$
\Gamma^{{\rm fin}}_\s(\fh):=\{ \Psi=(\Psi^{(0)},\cdots,\Psi^{(n)},\cdots) \in\Gamma_\s(\fh) \ | \ \Psi^{(n)}=0 \ {\rm for \ all \ but \ a \ finite \ number \ of \ }n\}, 
$$
the finite particle space. Note that $\Gamma^{{\rm fin}}_\s(\fh)$ is dense in $\Gamma_\s(\fh).$

For any $f\in \fh,$ $a(f)$ and $a^*(f)$ denote the usual annihilation/creation operators on $\Gamma_\s(\fh).$ They satisfy 
\beq\label{eq:ccr1}
[a(f_1),a(f_2)]=[a^*(f_1),a^*(f_2)]=0, \qquad [a(f_1),a^*(f_2)]=\langle f_1 | f_2\rangle .
\eeq
We also denote denote by $\phi(f):=\frac{1}{\sqrt{2}}(a(f)+a^*(f))$ the field operators and by $W(f):= \e^{\i\phi(f)}$ the Weyl operators. The Weyl operators are unitary and satisfy the following version of the CCR:
\beq\label{eq:ccrweyl}
W(f)W(g)=\e^{-\frac{\i}{2}\Im \langle f|g\rangle}W(f+g).
\eeq

If $h$ is an operator on $\fh,$ $\d\Gamma(h)$ will denote the second quantization of $h:$
$$
\d\Gamma(h)\lceil_{\otimes^n_\s \fh}:= \sum_{j=1}^n \underbrace{1\otimes \cdots \otimes 1}_{j-1}\otimes h\otimes \underbrace{1\otimes \cdots \otimes 1}_{n-j}.
$$
The operator $N:=\d\Gamma(1)$ is the number operator. The following estimates are well known and sometimes called $N_\tau-$estimates \cite{Ar,BFS,DJ,GJ}.

\bep\label{prop:abound} Let $h$ be a positive selfadjoint operator on $\fh,$ and $f\in \fh.$ Then, for all $\Psi \in \FDom (\d\Gamma(h)^{1/2}),$
\begin{eqnarray*}
\|a(f)\Psi\| & \leq & \|h^{-1/2}f\| \| \d\Gamma(h)^{1/2}\Psi\|,\\ 
\|a^*(f)\Psi\| & \leq & \|h^{-1/2}f\| \| (1+\d\Gamma(h))^{1/2}\Psi\|.
\end{eqnarray*}
\eep

\noindent Finally, if $q$ is a bounded operator on $\fh,$ we define
$\Gamma(q) : \Gamma_\s(\fh) \to \Gamma_\s(\fh)$ by 
$
\Gamma(q)\lceil_{\otimes^n_\s \cH} := q\otimes \cdots \otimes q. 
$


\subsection{Quadratic annihilation and creation operators}\label{ssec:doubleaa*}

Let $v \in \Gamma_\s^2(\fh).$ We define the annihilation and creation operators associated to $v$ as follows:
$$
a^*(v)\Psi:=\sqrt{n+2}\sqrt{n+1}v\otimes_\s\Psi, \qquad \Psi\in\Gamma_\s^n(\fh),
$$
$$
a(v)\Psi:=\sqrt{n+2}\sqrt{n+1} \left(\langle v|\, \otimes1^{\otimes n}\right)\Psi, \qquad \Psi\in\Gamma_\s^{n+2}(\fh),
$$
where $\langle v|\,\otimes1^{\otimes n}:\Gamma_\s^{n+2}(\fh)\ni
f_1\otimes \cdots \otimes f_{n+2}\mapsto \langle v| f_1\otimes f_2\rangle f_3\otimes \cdots \otimes f_{n+2}\in \Gamma_\s^n(\fh).$
These operators are well defined on $\Gamma^{{\rm fin}}_\s(\fh)$ and can be extended to $\FDom(N).$ 
\bep\label{prop:aa*dom} Let $\Psi \in \Gamma^{{\rm fin}}_\s(\fh),$ then 
\beq\label{eq:abound}(i)\quad \|a(v) \Psi\|  \leq  \|v\| \|N\Psi\|, \hspace{100mm}\eeq
\beq\label{eq:a*bound}(ii)\quad \|a^*(v) \Psi\|  \leq  \|v\| \|(N+2)^{1/2}(N+1)^{1/2}\Psi\|. \hspace{70mm}\eeq
\eep

\noindent This result will be a particular case of Propositions \ref{prop:arelbound} and \ref{prop:a*relbound} (Section \ref{sssec:sarelbound}).

Note also that if we write $v= \sum \lambda_n \ \phi_n \otimes \psi_n$, where
$(\phi_n)_{n\in\nn},(\psi_n)_{n\in\nn}$ are two orthonormal bases of
$\fh$ and $(\lambda_n)_{n\in\nn}$ is a sequence of positive numbers (with
$\sum \lambda_n^2=\|v\|_{\Gamma_\s(\fh)}^2<+\infty$), then we have
\beq\label{eq:defdoubleaa*}
a(v)= \sum \lambda_n a(\phi_n) a(\psi_n), \quad a^*(v)= \sum \lambda_n a^*(\phi_n) a^*(\psi_n),
\eeq
where on the right-hand side $a$ and $a^*$ denote the usual
annihilation/creation operators.

Before going further, we would like to make the link between elements
of the 2-particle space and real symmetric Hilbert-Schmidt operators
on $\fh,$ which will play an important role in the sequel. Let us fix
a conjugation $f\mapsto\bar{f}$ on $\fh$.
We denote
by $B^2(\fh)$ the set of all Hilbert-Schmidt operators and by
$B^2_\s(\fh)$ the set of real symmetric (i.e. $\bar{v}=v^*$)
Hilbert-Schmidt operators. It is well known that $\fh\otimes \fh$ and
$B^2(\fh)$ are isomorphic (the map
$
T: \fh\otimes \fh \ni \phi\otimes \psi \mapsto |\phi\rangle \langle \bar{\psi}| \in B^2(\fh)
$
extends by linearity and defines an isometry). It is easy to see that $T(\Gamma_\s^2(\fh))=B^2_\s(\fh).$ 
We will thus make no difference between a symmetric
Hilbert-Schmidt operator and the corresponding element of $\Gamma_\s^2(\fh).$

Using (\ref{eq:ccr1}), one then easily gets the following commutation relations:

\bep For all $v,v' \in \Gamma_\s^2(\fh),$ $f\in\fh$ and $h$ selfadjoint operator on $\fh,$
\beq\label{ccr2}
[a^*(v),a(f)]=-2a^*(v \bar{f}), \quad [a(v),a^*(f)]=2a(v \bar{f}),
\eeq
\beq\label{ccr3}
[a(v),a^*(v')]=4\d \Gamma(v'v^*) + 2 \Tr (v^* v').
\eeq
\beq\label{ccr4}
[\d\Gamma(h),a^*(v)]=a^*(hv+v\bar{h}), \quad [\d\Gamma(h),a(v)]=-a(hv+v\bar{h})
\eeq
\eep

To end this section, we would like to introduce the exponential of the operators $a(v)$ and $a^*(v),$ which will be used to define the unitary operators $U_\nat$ (Section \ref{ssec:symplmap}). 

\bep\label{prop:expaa*} Let $v\in B_\s^2(\fh).$  
\begin{itemize}
\item [1)] For all $\Psi\in\Gamma^{{\rm fin}}_\s(\fh),$ there exists
$\displaystyle{
\slim_{n\to+\infty} \sum_{k=0}^n
\frac{1}{k!}\left(\frac{1}{2}a(v)\right)^k\Psi}=:\e^{\frac{1}{2}a(v)}\Psi,$ and $\e^{\frac{1}{2}a(v)}\Psi 
\in \Gamma^{{\rm fin}}_\s(\fh)$.
\item [2)] If $\|v\|_{B(\fh)}<1,$ then for all
      $\Psi\in\Gamma^{{\rm fin}}_\s(\fh),$ there exists
$\displaystyle{
\slim_{n\to+\infty} \sum_{k=0}^n \frac{1}{k!}\left(\frac{1}{2}a^*(v)\right)^k\Psi}=:\e^{\frac{1}{2}a^*(v)}\Psi.$ 
\end{itemize}
\eep

\proof The proof of $1)$ is obvious since the right-hand side reduces to a finite sum. Now, part $2)$ follows from the fact that the function
$$
f(z):= \sum_{n=0}^{\infty} \left( \frac{1}{2^nn!}\right)^2 \|a^*(v)^n\Omega\|^2 z^n
$$
converges for all $|z|<\frac{1}{\|v\|_{B(\fh)}^2}$ (see \cite{Ru2}). Indeed, it is sufficient to prove the proposition for vectors of the form $\Psi=a^*(f_1)\cdots a^*(f_m)\Omega,$ where $f_1,\cdots f_m \in\fh$. Now, we have, for all $n\in\nn,$
\begin{eqnarray*}
\|\sum_{k=0}^n  \frac{1}{k!}\left(\frac{1}{2}a^*(v)\right)^k\Psi\|^2
& = & \sum_{k=0}^n \|\frac{1}{k!}\left(\frac{1}{2}a^*(v)\right)^k\Psi\|^2 \nonumber \\
 & \leq & \|f_1\|^2\cdots \|f_m\|^2 \sum_{k=0}^{n} \left( \frac{1}{2^kk!}\right)^2 (2k+m+1)(2k+m)\cdots(2k+1)
\|a^*(v)^k\Omega\|^2 \nonumber \\
 & \leq & \|f_1\|^2\cdots \|f_m\|^2 \sum_{k=0}^{+\infty}
 \frac{(2k+m+1)!}{(2k)!} \left( \frac{1}{2^kk!}\right)^2
 \|a^*(v)^k\Omega\|^2=:C_m\|\Psi\|^2<+\infty.
\end{eqnarray*}
\hfill\qed

\noindent Finally, we have the following
\bep\label{prop:expa*cvg} Let $(v_l)_{l\in\nn}$ be a sequence in $\Gamma_\s^2(\fh)$ such that
$\|v_l\|_{B(\fh)}<1$ for all $l\in\nn$ and $\lim_{l\to\infty} \|v_l\|=0$. Then the
operators $\e^{\frac{1}{2}a^*(v_l)}$ strongly converge to the identity on $\Gamma_\s^\fin(\fh)$.
\eep

\proof The result follows by the same computation as in the proof of
the previous proposition. Indeed, let $\Psi=a^*(f_1)\cdots
a^*(f_m)\Omega$, then for all $n\in\nn$, and using (\ref{eq:a*bound}),
$$
\|\sum_{k=0}^n
\frac{1}{k!}\left(\frac{1}{2}a^*(v_l)\right)^k\Psi-\Psi\|^2 \leq \|f_1\|^2\cdots \|f_m\|^2 \sum_{k=1}^{+\infty}
 \frac{(2k+m+1)!}{(2k)!} \left( \frac{1}{2^kk!}\right)^2
 \|a^*(v_l)^k\Omega\|^2\leq C_m \|v_l\| \|\Psi\|^2,
$$
where $C_m<+\infty.$ Hence
$\|\e^{\frac{1}{2}a^*(v_l)}\Psi-\Psi\|\leq  C_m \|v_l\|
\|\Psi\|^2$ and the result follows.
\hfill\qed


\subsection{Fock representations of CCR}\label{ssec:fockrepr}

In this paper we are interested in Fock representations of CCR, {\emph i.e.} $\cH=\Gamma_\s(\fh)$ where $\fh$ is a given complex Hilbert space. From now on, we assume that the real symplectic space $\cY$ is of the form $\cY=\{ (f,\bar{f}) \in \fh\oplus\fh | f\in\fh\}$ and that the symplectic form $\sigma ((f,\bar{f}),(g,\bar{g}))=\Im \langle f|g\rangle,$ where $\langle \cdot |\cdot\rangle$ denotes the scalar product in $\fh$.

We consider the map $\cY \ni (f,\bar{f})\mapsto W(f) \in U(\Gamma_\s(\fh))$
where $W(f)$ is the Weyl operator defined in Section \ref{ssec:generfock}. Using (\ref{eq:ccrweyl}), we can see that this map is a representation of CCR. Moreover, it is well known that this representation is regular and irreducible \cite{BR}.

Finally, we define the following operator on $\cY:$ $J=\left(\begin{array}{cc} \i & 0 \\ 0 & -\i \end{array}\right).$ This operator is an antiinvolution ($J^2=-1$) which preserves the symplectic form $\sigma.$


\section{Bogoliubov transformations and Bogoliubov Hamiltonians}\label{sec:dynsyst}


\subsection{Bogoliubov implementer}\label{ssec:symplmap}

A bounded real map $R$ on $\cY=\{ (f,\bar{f}) | f\in\fh\}$ will be written as
$
R= \left( \begin{array}{cc} P & \bar{Q} \\ Q & \bar{P} \end{array}\right),
$
where $P$ and $Q$ are bounded linear maps on $\fh,$ and $\bar{P}f:=\overline{P\bar{f}}$ (and similarly for $\bar{Q}$).
It is easy to see that a map $R$ is symplectic if and only if $RJR^*=R^*JR=J$ which is equivalent to 
\begin{equation}\label{eq:symplcond}
\left\{\begin{array}{rclrcl}
P^*P-Q^*Q & = & 1, & PP^*-\overline{QQ^*} & = & 1,\\
\bar{P^*}Q-\bar{Q^*}P & = & 0, & QP^*-\overline{PQ^*} & = & 0.
\end{array}\right.
\end{equation}
In particular, if $R$ is symplectic, then $P^*P\geq 1$ and therefore $P$ is invertible.

The following natural identification will be sometimes useful
\beq\label{def:identif}
I: \fh \ni f \mapsto (f,\bar{f})\in \cY.
\eeq
Given a symplectic map $R,$ we define
$W_R(f):= W(I^{-1}R(f,\bar{f})).$ 
The map $(f,\bar{f})\mapsto W_R(f)$ is also a
representation of CCR over $\cY$ in $\Gamma_\s(\fh)$.

\bed A symplectic map $R$ is called unitarily implementable if and
only if there exists a unitary operator $U$ on $\Gamma_\s(\fh)$ such
that $UW(f)U^{-1}=W_R(f),$ $\forall f\in\fh.$ If it exists, $U$ is
called a Bogoliubov implementer of $R$.
\eed

\bea\label{assu:shale} (Shale condition): $Q\in B^2(\fh)$
$(\Leftrightarrow [R,J]\in B^2(\fh\oplus\fh))$.  
\eea

We define the operators $K$ and $L$ as follows
\beq\label{def:operkl}
K:= \overline{QP^{-1}}, \qquad   L:=-P^{-1}\bar{Q}.
\eeq 

The following result is well known (see \cite{Be,Ru1,Ru2,Sh}).
\bet\label{thm:natimplem} $R$ is unitarily implementable if and only
if the Shale condition is satisfied. If it is satisfied, then
\begin{itemize}
\item[(i)] the operators $K$ and $L$ belong to $B^2_\s(\fh)$ and $\|K\|<1,$  
\item[(ii)] the operator 
\beq\label{def:u(r)}
U_\nat:= {\rm det} (1-K^*K)^{1/4} \e^{-\frac{1}{2}a^*(K)}\Gamma((P^{-1})^*) \e^{-\frac{1}{2}a(L)}
\eeq
is well defined on $\Gamma_\s^{{\rm fin}}(\fh),$ extends to a unitary
operator on $\Gamma_\s(\fh)$, and implements $R$.  
\end{itemize} 
\eet
\noindent We call $U_\nat$ the natural Bogoliubov implementer of $R$.
Since the Weyl representation is irreducible, if $R$ is
unitarily implementable, then the Bogoliubov implementer is unique up
to a phase factor. $U_\nat$ has the particular feature that its
expectation value on the vacuum is positive: $\langle
\Omega|U_\nat \Omega\rangle = {\rm det} (1-K^*K)^{1/4}>0.$


\subsection{Bogoliubov dynamics and Bogoliubov Hamiltonians}\label{ssec:qdyn}

Suppose $t\mapsto R(t)=\left(\begin{array}{cc} P(t) & \bar{Q}(t) \\ Q(t) &
    \bar{P}(t) \end{array}\right)$ is a strongly continuous one
parameter group of symplectic maps. We denote by $K(t)$ and $L(t)$ the operators
defined in (\ref{def:operkl}) associated to $R(t)$.

\bed\label{def:unitimpl} A one parameter symplectic
group $R(t)$ is called unitarily implementable if and only if there exists a
strongly continuous unitary group $U(t)$ such that, for all $t$, $U(t)$
is a Bogoliubov implementer of $R(t)$. If $R(t)$ is unitarily implementable, 
we call a Bogoliubov dynamics implementing $R(t)$ the unitary group
$U(t)$ and a Bogoliubov Hamiltonian (associated to $R(t)$) its selfadjoint generator.
\eed

Since the Bogoliubov implementer of a symplectic map $R$ is unique up
to a phase, if $R(t)$ is unitarily implementable, then there exists
$c(t)\in\cc,$ $|c(t)|=1$, such that $U(t)=c(t)U_\nat(t)$, and where
$U_\nat(t)$ is the natural Bogoliubov implementer of $R(t).$ $c(t)$ will be
called the natural cocycle for $U(t)$.

One can actually prove that $R(t)$ is unitarily implementable under very weak assumptions. 

\bet\label{thm:abstunitimpl} Suppose $R(t)$ is a strongly continuous
one-parameter symplectic group. Then $R(t)$ is unitarily implementable
if and only if the Shale condition is satisfied for all time $t$ and
$\lim_{t\to 0} \|K(t)\|_2=0$.
\eet

\proof Suppose $R(t)$ is unitarily implementable. Using Theorem
\ref{thm:natimplem}, we immediately get that the Shale condition is
satisfied for all $t$. It remains to prove that $\|K(t)\|_2\to 0$ as
$t$ goes to zero. Let $U(t)$ be a strongly continuous unitary group implementing
$R(t)$ and let 
\[
\alpha_t: \cB(\Gamma_\s(\fh)) \ni B \mapsto U(t)BU(t)^*\in\cB(\Gamma_\s(\fh)).
\]
Clearly $\alpha_t$ is a weak$*$ continuous one parameter group of $*-$automorphisms,
and $\alpha_t(B)=U_\nat(t)BU_\nat(t)^*$ since we
have $U(t)=c(t)U_\nat(t)$ where $c(t)$ is the natural cocycle for $U(t)$.
Therefore the map
$$
\rr\ni t\mapsto \Tr(|\Omega\rangle\langle \Omega | \alpha_t(|\Omega\rangle\langle \Omega |))=\det(1-K(t)^*K(t))^{1/2}
$$
is continuous. Since $\|K(t)\|<1,$ 
$\det(1-K(t)^*K(t))=\e^{\Tr(\log(1-K(t)^*K(t)))}$. Moreover $K(0)=0$,
so
$$
\lim_{t\to 0} \Tr(\log(1-K(t)^*K(t)))=0,
$$ 
from which the result follows using 
$$
\|K(t)\|_2^2=\Tr(K(t)^*K(t))\leq |\Tr(\log(1-K(t)^*K(t)))|.
$$

Suppose now that Shale condition is satisfied for
all $t$. Hence, for all $t$, we can construct $U_\nat(t)$ the natural
implementer associated to $R(t)$. Let us define the map
\[
\alpha_t: \cB(\Gamma_\s(\fh)) \ni B \mapsto U_\nat(t)BU_\nat(t)^*\in\cB(\Gamma_\s(\fh)).
\]
Obviously, for all $t,$ $\alpha_t$ is a weak$*$ continuous
$*-$automorphism of $\cB(\Gamma_\s(\fh)).$ Moreover, for all $t,s$
$\in\rr$,
$$
\alpha_t(\alpha_s(W(f)))=\alpha_{t+s}(W(f))=W_{R(t+s)}(f), \quad
\forall f\in\fh.
$$
Since the $*-$algebra generated by the Weyl operators is weak$*$ dense
in $\cB(\Gamma_\s(\fh)),$ this proves that $\alpha_t$ forms a
one-parameter group of $*-$automorphisms of $\cB(\Gamma_\s(\fh)).$ 

In order to prove that it can be implemented by a selfadjoint operator
$H$, it remains to show that this one parameter group is weak$*$
continuous with respect to $t$ (\cite{BR}, Ex 3.2.35). Moreover, using
the group property it suffices to prove that it is weak$*$ continuous
at $t=0.$ For that purpose, we shall prove that $t\mapsto U_\nat(t)$ is strongly
continuous at $t=0$.

The map $t\mapsto K(t)$ is continuous at $t=0$ in the
Hilbert-Schmidt norm by assumption (recall that $K(0)=0$). This together with Proposition
\ref{prop:expa*cvg} proves that $t\mapsto U_\nat(t)\Omega$ is
continuous at $t=0$.

Now, for any $f\in\fh$ one has
$U_\nat(t)W(f)\Omega=W_{R(t)}(f)U_\nat(t)\Omega.$ Hence
\begin{eqnarray*}
\| U_\nat(t)W(f)\Omega-W(f)\Omega\| & \leq & \|W_{R(t)}(f)(U_\nat(t)\Omega-\Omega)\|+ \| (W_{R(t)}(f)-W(f))\Omega\|\\
 & \leq & \|U_\nat(t)\Omega-\Omega\|+ \| (W_{R(t)}(f)-W(f))\Omega\|.
\end{eqnarray*}
The first term of the second line goes to zero as $t$ goes to zero,
and the second one as well since $t\mapsto R(t)$ is strongly
continuous and 
$$
\lim_{n\to+\infty} \|f_n-f\|=0 \Longrightarrow \slim_{n\to+\infty} W(f_n)=W(f).
$$
Thus, we have proven that $U_\nat(t)$ is strongly continuous at $t=0$
on ${\rm Span}\{W(f)\Omega,\, f\in\fh\}.$ Since this subspace is dense
in $\Gamma_\s(\fh)$ and the $U_\nat(t)$ are unitary, this ends the proof.
\hfill\qed


\subsection{Generator of unitarily implementable symplectic groups}\label{ssec:clasdyn}

In this section, we look for conditions on the generator $A$ of $R(t)$
which make it unitarily implementable. The basic assumption on the generator $A$ will be the following.

\bea\label{assu:gensympl} $A$ can be written as
$
A= \i\left( \begin{array}{cc} h & -v \\ \bar{v} & -\bar{h} \end{array}\right),
$
where $h$ is a selfadjoint operator with domain $\FDom(h),$ $v$ is a bounded operator such that $v^*=\bar{v},$ and $\FDom (A)= \FDom(h)\oplus \FDom(\bar{h}).$ 
\eea

\bep\label{prop:gensympl} Suppose $A$ satisfies Assumption \ref{assu:gensympl}, then $A$ generates a strongly continuous one-parameter group $(R(t))_{t\in\rr}$ of symplectic maps.
\eep

\proof $h$ is selfadjoint, therefore the operator 
$A_0:=\i \left( \begin{array}{cc} h & 0 \\ 0 & -\bar{h} \end{array}\right)$
generates a one-parameter group of unitary maps
$R_0(t)=\e^{t A_0}$.
Moreover, one can see that $R_0(t)$ is symplectic. Let us also write 
$V:=\i\left( \begin{array}{cc} 0 & -v \\ \bar{v} & 0 \end{array}\right).$
Then $A=A_0+V$ where $A_0$ is the generator of a strongly continuous one-parameter group and $V$ is a bounded operator. Hence, $A$ generates a one parameter strongly continuous group $R(t)$. 

Since $R(t)$ and $J$ leave $\FDom (A)$ invariant, for all $f\in \FDom(A),$ $t\mapsto R(t)JR(t)^*f$ is differentiable, and 
\[
\frac{\d}{\d t} R(t)JR(t)^*f=R(t)(JA^*+AJ)R(t)f.
\]
But, using $h^*=h$ and $v^*=\bar{v},$ one gets $(JA^*+AJ)f=0$ for all $f\in \FDom (A).$
Hence, $R(t)JR(t)^*=J$ on $\FDom (A).$ Since they are both bounded
operators and $\FDom(A)$ is dense, this proves that $R(t)JR(t)^*=J$ on
$\fh\oplus \fh.$ We prove similarly that $R(t)^*JR(t)=J,$ so that
$R(t)$ is symplectic. 
\hfill\qed

From now on, we will always assume that Assumption \ref{assu:gensympl} is satisfied.
Let us define
\beq\label{def:Taverage}
v(t):=\int_0^t \e^{\i\tau h}v\e^{\i\tau \bar{h}}\d\tau.
\eeq

\bea\label{assu:genshalemin} For all $t,$ $v(t)\in B^2(\fh)$, the
function $t\mapsto \|v(t)\|_2$ is locally integrable on $\rr$ and continuous at
$t=0$. 
\eea

\noindent This assumption was already used in \cite{Be}. 

\bet\label{thm:genshale}  Suppose $A$ satisfies Assumption \ref{assu:genshalemin}. Then $R(t)$ is unitarily implementable.
\eet 

\proof We define $V(t):= R_0(t)VR_0(-t)$ and $\tilde{R}(t):=R(t)R_0(-t).$ Since $V$ is bounded, we have
\beq\label{eq:rintegral}
\tilde{R}(t)=1+\int_0^t \tilde{R}(\tau)V(\tau) \d\tau.
\eeq
We introduce the following sequence of bounded operators
$$
\tilde{R}_0(t) = 1, \qquad \tilde{R}_{n+1}(t) = \int_0^t \tilde{R}_n(\tau)V(\tau) \d\tau.
$$
In particular we have 
\[
\tilde{R}_1(t)=\int_0^t V(\tau) \d\tau= \i\left( \begin{array}{cc} 0 & -v(t) \\ \overline{v(t)} & 0 \end{array}\right).
\]
Hence, $\tilde{R}_1(t)$ is Hilbert-Schmidt and $\|\tilde{R}_1(t)\|_2 = \sqrt{2}\|v(t)\|_2.$
Then, using $\|V(\tau)\|=2\|v\|$ for all $\tau,$ we get
\begin{eqnarray*}
\|\tilde{R}_{n+1}(t)\|_2 & \leq & 2\|v\| \int_0^t \|\tilde{R}_n(\tau)\|_2 \d\tau\\
 & \leq & (2\|v\|)^n \int_0^t \frac{(t-\tau)^{n-1}}{(n-1)!}\sqrt{2}\|v(\tau)\|_2\d\tau, \qquad \forall n\geq 1.
\end{eqnarray*}
Moreover, we have $\tilde{R}(t)-1=\sum_{n\geq 1} \tilde{R}_n(t),$ hence
\beq\label{eq:rbound}
\|\tilde{R}(t)-1\|_2 \leq \sqrt{2}\|v(t)\|_2 + 2\sqrt{2}\|v\|\int_0^t \e^{2(t-\tau)\|v\|} \|v(\tau)\|_2 \d\tau <+\infty.
\eeq
Since $R(t)R_0(-t)-1$ is Hilbert-Schmidt, so is $R(t)-R_0(t).$ Now,
$Q_0(t)=0$ hence $Q(t)$ is Hilbert-Schmidt. 

It remains to prove that $\lim_{t\to 0} \|K(t)\|_2=0.$ Using
(\ref{eq:rbound}) and the continuity of $\|v(t)\|_2$ at $t=0$, we get
$\lim_{t\to 0} \|\tilde{R}(t)-1\|_2=0.$
Thus we have $\lim_{t\to 0} \|\bar{Q}(t)\e^{\i t\bar{h}}\|_2=0.$
Finally, by definition of $K(t)$ we have
$$
\qquad\qquad\qquad\qquad\|K(t)\|_2 = \| \overline{Q(t)P(t)^{-1}}\|_2
\, \leq \, \|\bar{Q}(t)\e^{\i t\bar{h}}\|_2 \|\e^{-\i
   t\bar{h}}\overline{P(t)^{-1}}\| \, \leq  \, \|\bar{Q}(t)\e^{\i t\bar{h}}\|_2.\qquad\qquad\qquad\Box
$$


\subsection{Bogoliubov dynamics of type I}\label{ssec:dyntype1}

As mentioned in the introduction, there are natural choices for the
Bogoliubov dynamics implementing $R(t)$, one of them being type I.
However, it is not always possible to define it and one has
to impose some additional assumption on
$R(t)$. We will denote by $B^1(\fh)$ the set of trace class operators on $\fh$
and by $\|\cdot\|_1$ the trace norm.

\bed Let $t\mapsto R(t)$ be a unitarily implementable symplectic group,
with generator $A$. We say that it is type I
if and only if, for all $t\in\rr$, $P(t)\e^{-\i th}-1\in B^1(\fh)$ 
and $\lim_{t\to 0} \|P(t)\e^{-\i th}-1 \|_1=0$.
\eed

\bet\label{thm:utildegroup} Suppose $R(t)$ is a type I symplectic group. Then, the operators
\begin{equation}\label{def:utilde}
U_I(t) := {\rm det}(\bar{P}(t)\e^{\i t\bar{h}})^{-1/2}\e^{-\frac{1}{2}a^*(K(t))}\Gamma((P(t)^{-1})^*) \e^{-\frac{1}{2}a(L(t))} 
\end{equation}
form a Bogoliubov dynamics implementing $R(t)$. Their natural cocycle 
is given by
\begin{equation}\label{def:natcocycle}
c_I(t) = {\rm det}(\bar{P}(t)\e^{\i t\bar{h}})^{-1/2} {\rm det}(1-K(t)^*K(t))^{-1/4}. 
\end{equation}
\eet

\bed The operator $H_I=\frac{1}{\i}\frac{\d}{\d
  t}U_I(t)\lceil_{t=0}$ is called a Bogoliubov Hamiltonian of type I.
\eed

\noindent In the proof of Theorem \ref{thm:utildegroup}, we will make use of the following lemmas.

\bel\label{lem:detcommut} Let $B$ be a bounded operator and $V$ a unitary operator such that $BV-1$ is trace class. Then $VB-1$ is trace class and ${\rm det}(BV)={\rm det}(VB).$ 
\eel

\bel\label{lem:exp/det} Let $K,$ $L\in B^2(\fh)$ such that $\bar{K}=K^*, \bar{L}=L^*$ and $\|K\|<1,\|L\|<1.$ Then 
\[
\langle \e^{-\frac{1}{2}a^*(L)\Omega}|\e^{-\frac{1}{2}a^*(K)}\Omega\rangle = {\rm det}(1-L^*K)^{-1/2}.
\]
\eel

\proof Since $K$ is Hilbert-Schmidt and $\bar{K}=K^*,$ there exist an orthonormal basis of $\fh,$ $(f_n)_n,$ and a sequence $\lambda_n$ such that $K=\sum \lambda_n |f_n\rangle\langle \bar{f}_n|.$ Similarly, we can write $L= \sum \mu_m  |g_m\rangle\langle \bar{g}_m|.$ Therefore, we have
\begin{eqnarray*}
\langle \e^{-\frac{1}{2}a^*(L)}\Omega|\e^{-\frac{1}{2}a^*(K)}\Omega\rangle & = & \prod_{m,n} \langle \e^{-\frac{1}{2}\mu_ma^*(g_m)^2}\Omega|\e^{-\frac{1}{2}\lambda_na^*(f_n)^2}\Omega\rangle\\
 & = & \prod_{m,n} \sum_j \left( -\frac{1}{2}\right)^{2j}\frac{\bar{\mu}_m^j\lambda_n^j(2j)!}{(j!)^2}\langle g_m|f_n \rangle^{2j}\\
 & = & \prod_{m,n} (1-\bar{\mu}_m\lambda_n\langle g_m|f_n \rangle^2)^{-1/2}.
\end{eqnarray*}
Now, it suffices to see that $L^*K= \displaystyle{\sum_{m,n}} \bar{\mu}_m\lambda_n \langle g_m|f_n\rangle |\bar{g}_m\rangle\langle\bar{f}_n|.$
\hfill\qed

\noindent {\bf Proof of Theorem \ref{thm:utildegroup}}\ \ Since
$U_\nat(t)$ is unitary and $U_I(t)=c_I(t)U_\nat(t)$, to prove that
$U_I(t)$ is a Bogoliubov implementer, it suffices to show that
$|c_I(t)|=1.$
Using (\ref{eq:symplcond}), we have $1-K(t)^*K(t)=(\overline{P(t)P(t)^*})^{-1}.$ Now
\begin{eqnarray*}
|{\rm det}(\bar{P}(t)\e^{\i t\bar{h}})^{-1/2}| = \left| {\rm
    det}(\bar{P}(t) \e^{\i t\bar{h}}) \overline{{\rm
      det}((\bar{P}(t)\e^{\i t\bar{h}})^*)} \right|^{-1/4} & = &
\left| {\rm det}(\bar{P}(t)\e^{\i t\bar{h}}) {\rm det}(\e^{-\i t\bar{h}}\bar{P(t)^*}) \right|^{-1/4}\\
 & = & |{\rm det}(\overline{P(t)P(t)^*})|^{-1/4}.
\end{eqnarray*}
We now prove that the operators $U_I(t)$ form a one-parameter group. As for $U_\nat(t),$ for all $s$ and $t$ there exists $\alpha(t,s)\in\rr$ such that 
$U_I(t)U_I(s)=\e^{\i\alpha(t,s)}U_I(t+s).$
Using Lemmas \ref{lem:detcommut} and \ref{lem:exp/det} we have 
\begin{eqnarray*}
\langle \Omega|U_I(t)U_I(s)\Omega\rangle & = & {\rm
  det}(\bar{P}(t)\e^{\i t\bar{h}})^{-1/2}{\rm det}(\bar{P}(s)\e^{\i s\bar{h}})^{-1/2} \langle \e^{-\frac{1}{2}a^*(L(t))}\Omega|\e^{-\frac{1}{2}a^*(K(s))}\Omega\rangle\\
 & = & {\rm det}(\bar{P}(t)\e^{\i t\bar{h}})^{-1/2}{\rm
   det}(\bar{P}(s)\e^{\i s\bar{h}})^{-1/2} {\rm det}(1-L(t)^*K(s))^{-1/2}\\
 & = & \left( {\rm det}(\e^{\i t\bar{h}}\bar{P}(t)){\rm
     det}(1+\bar{P}(t)^{-1}Q(t)\bar{Q}(s)\bar{P}(s)^{-1}){\rm
     det}(\bar{P}(s)\e^{\i s\bar{h}}) \right)^{-1/2}\\
 & = & {\rm det} (\e^{\i
   t\bar{h}}(\bar{P}(t)\bar{P}(s)+Q(t)\bar{Q}(s))\e^{\i s\bar{h}})^{-1/2}\\
 & = & {\rm det} (\bar{P}(t+s)\e^{\i(t+s)\bar{h}})^{-1/2}=\langle \Omega |U_I(t+s)\Omega\rangle.
\end{eqnarray*}
Therefore $\e^{\i\alpha(t,s)}=1$ and $U_I(t)$ is a one-parameter group.

Finally we have to prove that $U_I(t)$ is strongly continuous. Using
the group property together with the same argument as in Theorem
\ref{thm:abstunitimpl}, it suffices to prove that $t\mapsto
U_I(t)\Omega$ is continuous at $t=0$. 
Now, $t\mapsto K(t)$ is
continuous in the Hilbert-Schmidt norm since $R(t)$ is unitarily
implementable (Theorem
\ref{thm:abstunitimpl}), and, by assumption, $t\mapsto P(t)\e^{-\i th}$
is continuous in the trace norm at $t=0$, thus so is the map $t\mapsto
\det(P(t)\e^{-\i th})$, which ends the proof.
\hfill\qed


\subsection{Generator of type I Bogoliubov dynamics}\label{ssec:typeIgen}

We would like in this section to give some sufficient conditions on the generator
$A$ of a symplectic group $R(t)$ so that it is of type I.

\bea\label{assu:gengroup} For all $t,$ the operator $\bar{v}v(t)$ is
trace class and the function $t\mapsto \|\bar{v}v(t)\|_1$ is locally
integrable on $\rr$ and continuous at $t=0$.
\eea

\noindent This condition was also used in \cite{Be}.

\bea\label{assu:genshale} $v$ is a Hilbert-Schmidt operator on $\fh$. 
\eea

\bet\label{thm:gentype1} 
\begin{itemize}
\item[(i)] If Assumptions \ref{assu:genshalemin} and
      \ref{assu:gengroup} are satisfied, then $R(t)$ is of type I.
\item[(ii)] If Assumption \ref{assu:genshale} is satisfied, then
      $R(t)$ is of type I and we have
\begin{equation}
U_I(t)= \e^{\frac{\i}{2} \Tr(\int_0^t Q(s)v\bar{P}(s)^{-1}\d s)}\e^{-\frac{1}{2}a^*(K(t))}\Gamma((P(t)^{-1})^*) \e^{-\frac{1}{2}a(L(t))}, \label{def:utilde2}
\end{equation}
\begin{equation}\label{def:natcocycle2}
c_I(t)= \e^{\frac{\i}{2} \Re(\Tr(\int_0^t Q(s)v\bar{P}(s)^{-1}\d s))}. 
\end{equation}
\end{itemize}
\eet 

\proof We will use the notation introduced in the proof of Theorem \ref{thm:genshale}. 
Let 
\[
\cV:=\left\{ R= \left( \begin{array}{cc} A & B \\ C & D \end{array}\right) \in \cB(\fh\oplus\fh) \,| \, A,D \in B^1(\fh), B,C\in B^2(\fh) \right\},
\]
with $\|R\|_\cV:= \|A\|_1+\|D\|_1+\|B\|_2+\|C\|_2.$ Note that if $R$ and $R'$ are in $\cV,$ then so is $RR'.$ 

$(i)$ Suppose Assumption \ref{assu:gengroup} is satisfied.
We have $\tilde{R}_1(t)\in\cV.$ Then, 
\[
\tilde{R}_2(t)=\int_0^t \tilde{R}_1(\tau) V(\tau) \d\tau= \left( \begin{array}{cc} \int_0^t v(\tau)\e^{-\i\tau \bar{h}}\bar{v}\e^{-\i\tau h}\d\tau & 0 \\ 0 & \int_0^t \overline{v(\tau)} \e^{\i\tau h}v\e^{\i\tau \bar{h}}\d\tau \end{array}\right).
\]
Using Assumption \ref{assu:gengroup}, one has 
\[
\|\int_0^t v(\tau) \e^{-\i\tau \bar{h}}\bar{v}\e^{-\i\tau h}\d\tau\|_1 \leq \int_0^t \|\bar{v}v(\tau)\|_1 \d\tau=:\rho(t)<+\infty
\]
Therefore $\tilde{R}_2(t)$ is trace class and $\|\tilde{R}_2(t)\|_1 \leq 2\rho(t).$
In the same way as in the proof of Theorem \ref{thm:genshale}, we have that $\tilde{R}(t)-1-\tilde{R}_1(t)$
is trace class, and hence is in $\cV.$ Thus so is $\tilde{R}(t)-1.$ In
particular, $P(t)\e^{-\i th}-1$ is trace class.

Finally, the continuity of $\|\bar{v}v(t)\|_1$ at $t=0$ implies the one
of $\|P(t)\e^{-\i th}-1\|_1$ in a similar way as in
Theorem \ref{thm:genshale}.

$(ii)$ Suppose now that Assumption \ref{assu:genshale} is
satisfied. First note that it implies Assumptions \ref{assu:genshalemin} and
 \ref{assu:gengroup}, so that $R(t)$ is of type I.
According to the definition of $U_I(t)$, we have to prove that, for
all $t,$ 
\begin{equation}\label{eq:detdiff}
{\rm det}(\bar{P}(t)\e^{\i t\bar{h}})=\e^{-\i\int_0^t \Tr(Q(s)v\bar{P}(s)^{-1})\d s}.
\end{equation}

For all $t,$ $V(t)\in\cV,$ and, using (\ref{eq:rintegral}), we have as an identity in $\cV$
$$
\tilde{R}(t)-1=\int_0^t V(\tau)\d\tau +\int_0^t (\tilde{R}(\tau)-1)V(\tau) \d\tau.
$$

We have $V(t)= \i\left( \begin{array}{cc} 0 & -\e^{\i
      th}v\e^{\i t\bar{h}} \\ \e^{-\i t\bar{h}}\bar{v}\e^{-\i th} & 0
  \end{array} \right).$ It is clear that $t\mapsto
\e^{\i th}v\e^{\i t\bar{h}}$ is continuous in the weak operator topology,
and therefore in the weak sense in $B^2(\fh)$ considered as a Hilbert
space (\emph{i.e.} for all $K\in B^2(\fh),$ $t\mapsto\Tr(K\e^{\i
  th}v\e^{\i t\bar{h}})$ is continuous). Moreover, since $\e^{\i th}$
is unitary, we have $\|\e^{\i th}v\e^{\i
  t\bar{h}}\|_2=\|v\|_2.$ But in a Hilbert space, a
function which takes values on a sphere and which is weakly continuous
is actually norm continuous. 
Hence $\e^{\i th}v\e^{\i t\bar{h}}$ is continuous in the
Hilbert-Schmidt norm. So $V(t)$ is
continuous in $\cV$ and thus $R(t)R_0(-t)-1$ is differentiable in $\cV$.
In particular, $\bar{P}(t)\e^{\i t\bar{h}}-1$ is
differentiable in the trace class topology. Hence, ${\rm
  det}(\bar{P}(t)\e^{\i t\bar{h}})$ is differentiable and 
\begin{eqnarray*}
\frac{\d}{\d t}{\rm det}(\bar{P}(t)\e^{\i t\bar{h}}) & = & \Tr
\left(\frac{\d}{\d t}(\bar{P}(t)\e^{\i
    t\bar{h}})\times(\bar{P}(t)\e^{\i t\bar{h}})^{-1}\right)\times{\rm
  det}(\bar{P}(t)\e^{\i t\bar{h}})\\ 
 & = & -\i\Tr (Q(t)v\bar{P}(t)^{-1})\times{\rm
   det}(\bar{P}(t)\e^{\i t\bar{h}}),
\end{eqnarray*}
which proves (\ref{eq:detdiff}), and where we used (\ref{eq:rintegral}) in the second line.  

The proof of (\ref{def:natcocycle2}) follows from
(\ref{def:natcocycle}), (\ref{eq:detdiff}) and the fact that ${\rm det}(1-K(t)^*K(t))$ is positive.
\hfill\qed


\subsection{Essential selfadjointness of type I Bogoliubov Hamiltonians}\label{ssec:quadham}

Formally, it is easy to see that the Bogoliubov Hamiltonian of type I
is given by (\ref{intro:ham}) with $c=0$. 
We can make this precise when $v$ is Hilbert-Schmidt.

\bet\label{thm:selfadj2} Suppose Assumption \ref{assu:genshale} is satisfied, then the
operator $H_I=\d\Gamma(h)+\frac{1}{2}(a^*(v)+a(v))$ is essentially
 selfadjoint on $\cD:=\Gamma_\s^{{\rm fin}}(\fh)\cap
 \FDom(\d\Gamma(h))$ and $\e^{\i t H_I}=U_I(t).$
\eet

Note that, since $\Gamma_\s^{{\rm fin}}(\fh) \subset \FDom(a^*(v)+a(v))$, the operator $H$ is
therefore essentially selfadjoint on $\FDom(\d\Gamma(h))\cap
\FDom(a^*(v)+a(v)).$ The strategy of the proof for the essential selfadjointness comes from \cite{Be} and goes back to Carleman \cite{Ca}. However, as we mentioned in the introduction, the proof in \cite{Be} is not completely rigorous. 
In the case where $h$ is bounded, a similar result has also been
proven in \cite{IH}.

Note also that $\Omega\in\cD(H_I)$ and that $H_I$ has the particular
feature that $\langle \Omega|H_I\Omega\rangle=0.$

Recall that $L(t)=-P(t)^{-1}\bar{Q}(t).$ When $v$ is Hilbert-Schmidt, the operator $v\bar{L}(t)$ is trace class and 
$\Tr (Q(s)v\bar{P}(s)^{-1})= -\Tr(v\bar{L}(s)).$  
Therefore,
\beq\label{eq:uutildephase}
c_I(t)=\e^{-\frac{\i}{2}\Re(\int_0^t \Tr(v\bar{L}(s))\d s)}.
\eeq

\bel\label{lem:Ltdiff} Suppose Assumption \ref{assu:genshale} is satisfied. Then the map $t\mapsto L(t)$ is differentiable in the Hilbert-Schmidt topology.
\eel

\proof In the same way as in Theorem \ref{thm:genshale}, we can prove
that $R_0(-t)R(t)-1$ is differentiable in $\cV.$ 
Hence, $\e^{-\i th}P(t) -1$ is differentiable in the trace class norm,
thus $\e^{-\i th}P(t)$ is norm differentiable and hence so is $P(t)^{-1}\e^{\i th}.$ Moreover $\e^{-\i th}\bar{Q}(t)$ is differentiable in
the Hilbert-Schmidt norm, so that $L(t)=-P(t)^{-1}\bar{Q}(t)=-(\e^{-\i
  th}P(t))^{-1}\e^{-\i th}\bar{Q}(t)$ is differentiable in the Hilbert-Schmidt norm.
\hfill\qed

\bel\label{prop:u(r)w-diff} Suppose Assumption \ref{assu:genshale} is satisfied, then $\langle \Omega| U_\nat(t) \Omega\rangle$ is continuously differentiable. 
\eel

\proof Using (\ref{def:natcocycle}) and (\ref{def:natcocycle2}) we
have $\langle \Omega| U_\nat(t) \Omega\rangle = \e^{\frac{1}{2}\Im(\int_0^t \Tr(v\bar{L}(s))\d s)}.$
The differentiability then follows from Lemma \ref{lem:Ltdiff}.
\hfill\qed

\noindent {\bf Proof of Theorem \ref{thm:selfadj2}}\ \ We first prove
that $H_I$ is essentially selfadjoint on $\cD$. For that purpose, we consider the
symmetric operator $H$ defined as $H_I$ on the domain $\cD$ and we prove that for all $z\in\cc, z\notin\rr,$ 
$\Ker(H^*-z)=\{0\}.$

We denote by $P_n$ the orthogonal projection onto $\Gamma_\s^n(\fh).$ In particular, for any vector $\Psi,$ $P_n\Psi\in\Gamma_\s^{{\rm fin}}(\fh).$ We also define, for all $\epsilon \in \rr,$ 
\[
\Psi_\epsilon:=\left(1-\i\epsilon\d\Gamma(h)\right)^{-1}\Psi.
\]
For any $\epsilon\neq 0,$ $\Psi_\epsilon \in \FDom(\d\Gamma(h))$ and $\lim_{\epsilon\to 0}\Psi_\epsilon =\Psi.$ Moreover, since the operator $\d\Gamma(h)$ leaves the subspace $\Gamma_\s^n(\fh)$ invariant, we have $P_n\Psi_\epsilon=(P_n\Psi)_\epsilon \in \cD$ for all $n$ and $\epsilon\neq 0.$

Let us now fix $z\notin\rr$ and let $\Phi\in \Ker(H^*-z).$ For all $n$ we have
$$
z\|P_n\Phi\|^2 = z \langle P_n\Phi|\Phi\rangle = \lim_{\epsilon\to 0} z\langle P_n\Phi_\epsilon|\Phi\rangle 
 = \lim_{\epsilon\to 0} \langle P_n\Phi_\epsilon|H^*\Phi\rangle 
 = \lim_{\epsilon\to 0} \langle HP_n\Phi_\epsilon|\Phi\rangle,
$$
where in the last equality we have used the fact that $P_n\Phi_\epsilon\in\cD.$ Similarly, we have
$\bar{z}\|P_n\Phi\|^2=\lim_{\epsilon\to 0} \langle \Phi|HP_n\Phi_{-\epsilon}\rangle.$
Therefore,
\begin{eqnarray*}
2\i\Im z\|P_n\Phi\|^2 & = & \lim_{\epsilon\to 0} ( \langle \d\Gamma(h)P_n\Phi_\epsilon|\Phi\rangle -\langle\Phi|\d\Gamma(h) P_n\Phi_{-\epsilon}\rangle +\frac{1}{2}\langle (a(v)+a^*(v)) P_n\Phi_\epsilon|\Phi\rangle\\
 & & \qquad \qquad \qquad \qquad \qquad \qquad \qquad \qquad \qquad \qquad \qquad -\frac{1}{2}\langle \Phi| (a(v)+a^*(v)) P_n\Phi_{-\epsilon}\rangle ).
\end{eqnarray*}
Since $P_n$ commutes with $\d\Gamma(h),$ 
the two first terms of the right hand side cancel. Moreover, the
operator $(a(v)+a^*(v)) P_n$ is bounded. So finally we get, with the convention $P_{-1}=P_{-2}=0,$
\begin{eqnarray*}
4\i\Im z\|P_n\Phi\|^2 & = & \langle (a(v)+a^*(v)) P_n\Phi|\Phi\rangle -\langle \Phi| (a(v)+a^*(v)) P_n\Phi \rangle\\
 & = & \langle a(v)P_n\Phi|P_{n-2}\Phi\rangle+\langle a^*(v)P_n\Phi|P_{n+2}\Phi\rangle -\langle a(v)P_{n+2}\Phi|P_n\Phi\rangle -\langle a^*(v)P_{n-2}\Phi|P_n\Phi\rangle.
\end{eqnarray*}

We now sum the previous identity for $0\leq n\leq N,$ which gives
\begin{eqnarray*}
4\i\Im z \sum_{n=0}^N\|P_n\Phi\|^2 & = & \langle a^*(v)P_N\Phi|P_{N+2}\Phi\rangle+\langle a^*(v)P_{N-1}\Phi|P_{N+1}\Phi\rangle -\langle a(v)P_{N+2}\Phi|P_N\Phi\rangle\\
 & & \qquad \qquad \qquad \qquad \qquad \qquad \qquad \qquad \qquad \qquad \qquad -\langle a(v)P_{N+1}\Phi|P_{N-1}\Phi\rangle.
\end{eqnarray*}
Therefore, for all $N\in\nn$, and using Proposition \ref{prop:aa*dom}, we have
\begin{eqnarray*}
4|\Im z| \sum_{n=0}^N\|P_n\Phi\|^2 & \leq & \|v\|_2\left( 2(N+2)\|P_N\Phi\| \|P_{N+2}\Phi\| +2(N+1)\|P_{N-1}\Phi\| \|P_{N+1}\Phi\|  \right)\\
 & \leq & (N+2)\|v\|_2 \left( \|P_{N-1}\Phi\|^2+ \|P_N\Phi\|^2+\|P_{N+1}\Phi\|^2+\|P_{N+2}\Phi\|^2 \right).
\end{eqnarray*}

Suppose $\Phi\neq 0.$ Hence there exists $N_0$ such that $\sum_{n=0}^{N_0}\|P_n\Phi\|^2=C>0,$ and for all $N\geq N_0,$ $\sum_{n=0}^N\|P_n\Phi\|^2\geq C.$ So we have, for all $N\geq N_0,$
\[
\frac{4|\Im z|C}{N+2} \leq \|v\|_2 \sum_{j=N-1}^{N+2} \|P_j\Phi\|^2.
\]
If now we sum over $N$ this inequality, the right hand side converges (and is less that $4\|v\|_2\|\Phi\|^2$), while the left hand side diverges. Hence $\Phi=0$ and $H_I$ is essentially selfadjoint on $\cD.$

It remains to prove that $\e^{\i tH_I}=U_I(t)$.
For that purpose, we prove that $\e^{\i tH_I}$ is a Bogoliubov
dynamics implementing $R(t)$ (first provided $h$ is bounded and then
for a general $h$) so that it equals $U_I(t)$ up to a phase
factor. And then we prove that this phase is $1$.

Given two operators $B$ and $C,$ let ${\rm ad}^0_BC:=C$ and ${\rm ad}^k_BC:=[B,{\rm ad}^{k-1}_BC].$ Recall also that $\phi(f)$ stand for the field operators on $\Gamma_\s(\fh).$

Suppose $h,$ and hence $A,$ is bounded. Then, for all $f\in\fh,$ and in
the sense of quadratic forms on $\Gamma_\s^{{\rm fin}}(\fh),$
\beq\label{eq:Hcommut}
{\rm ad}^k_{\i H_I}\phi(f)=\phi(I^{-1}A^k(f,\bar{f})),
\eeq
where $I$ was defined in (\ref{def:identif}).
Indeed, as quadratic forms on $\Gamma_\s^{{\rm fin}}(\fh),$ we have
\[
[\i H_I,a(f)] = \i[\d\Gamma(h),a(f)]+\frac{\i}{2}[a^*(v),a(f)] = a(\i hf) + a^*(
-\i v \bar{f}).
\]
In the same way, one proves that $[\i H_I,a^*(f)] = a^*(\i hf) + a( -\i
v \bar{f}).$
Hence, one has $[\i H_I,\phi(f)]=\phi(I^{-1}A(f,\bar{f})),$ 
and since $A$ is bounded (\ref{eq:Hcommut}) follows easily.

Let now $\Phi, \Psi \in \Gamma_\s^{{\rm fin}}(\fh).$ For $z\in \cc,$ we define
$$
F_1(z):= \langle \Phi, \e^{\i zH_I} \phi(f) \Psi\rangle \quad {\rm
  and} \quad
F_2(z):= \langle \phi(I^{-1}\e^{zA}(f,\bar{f}))\Phi, \e^{\i zH_I}\Psi\rangle.
$$
Using Proposition \ref{prop:aa*dom}, it is easy to see that $\Phi$ and
$\Psi$ are analytic for $H_I.$ Since moreover $A$ is bounded, this
proves that $F_1$ and $F_2$ are analytic functions in some neighborhood of $0.$ Moreover it is well known that 
$B^nC=\sum_{k=0}^n \left(\begin{array}{c} n \\ k \end{array}\right) {\rm ad}^k_BC\, B^{n-k}.$ 
Thus, for all $n$ we have
\begin{eqnarray*}
\frac{\d^nF_1(z)}{\d z^n}\lceil_{z=0} = \langle \Phi, (\i H_I)^n \phi(f) \Psi\rangle 
 & = & \sum_{k=0}^n \left(\begin{array}{c} n \\ k \end{array}\right)
 \langle \Phi, {\rm ad}^k_{\i H_I}\phi(f)\, (\i H_I)^{n-k}\Psi\rangle \\ 
 & = & \sum_{k=0}^n \left(\begin{array}{c} n \\ k \end{array}\right)
 \langle \Phi, \phi(I^{-1}A^k(f,\bar{f}))(\i H_I)^{n-k}\Psi\rangle \\
 & = & \frac{\d^nF_2(z)}{\d z^n}\lceil_{z=0},
\end{eqnarray*}
where we have used (\ref{eq:Hcommut}) in the second line.
Therefore, for all $z$ in some neighborhood of $0,$ $F_1(z)=F_2(z),$ which implies that 
\[
\e^{\i tH_I}\phi(f)\e^{-\i tH_I}\Phi=\phi(I^{-1}\e^{tA}(f,\bar{f}))\Phi,
\]
and hence
\[
\e^{\i tH_I}W(f)\e^{-\i tH_I}\Phi=W_{R(t)}(f)\Phi,
\]
for all $\Phi \in \Gamma_\s^{{\rm fin}}(\fh),$ and all $t\in \rr$ by
the group property. Since $\Gamma_\s^{{\rm fin}}(\fh)$ is dense in
$\Gamma_\s(\fh),$ this proves that $\e^{\i tH_I}$ intertwines $W$ and $W_{R(t)}$ when $h$ is bounded.

We consider now the general case. Let us write $A=A_0+V$ as in the
proof of Proposition \ref{prop:gensympl}.
Both $A_0$ and $V$ are generators of a one-parameter group of symplectic maps. Moreover, $V$ is bounded, therefore we can apply the first part of the proof and, for all $t\in \rr,$ we have
\[
\e^{\frac{\i}{2} t(a^*(v)+a(v))}W(f)\e^{-\frac{\i}{2} t(a^*(v)+a(v))}=W(I^{-1}\e^{tV}(f,\bar{f})).
\]
But, since $h$ is selfadjoint, it is well known (see \emph{e.g.} \cite{DG}) that 
\[
\e^{\i t\d\Gamma(h)}W(f)\e^{-\i t\d\Gamma(h)}=W(\e^{\i th}f)=W(I^{-1}\e^{tA_0}(f,\bar{f})).
\]
Thus, using the Trotter product formula,
$$
\e^{\i tH_I} = \slim_{n\to\infty} \left(\e^{\frac{\i
      t\d\Gamma(h)}{n}}\e^{\frac{\i t(a^*(v)+a(v))}{2n}} \right)^n
 = \slim_{n\to\infty} \left(\e^{\frac{\i
       t(a^*(v)+a(v))}{2n}}\e^{\frac{\i t\d\Gamma(h)}{n}} \right)^n.
$$
In the same way, we have
$R(t)=\slim_{n\to\infty} \left( \e^{\frac{tA_0}{n}}\e^{\frac{tV}{n}}\right)^n.$
Hence,
\begin{eqnarray*}
\e^{\i tH_I}W(f)\e^{-\i tH_I} & = & \slim_{n\to\infty} \left(\e^{\frac{\i
      t\d\Gamma(h)}{n}}\e^{\frac{\i t(a^*(v)+a(v))}{2n}}
\right)^nW(f)\left(\e^{\frac{-\i t(a^*(v)+a(v))}{2n}}\e^{\frac{-\i t\d\Gamma(h)}{n}} \right)^n \\
 & = & \slim_{n\to\infty} W\left(I^{-1} \left(
     \e^{\frac{tA_0}{n}}\e^{\frac{tV}{n}}\right)^n(f,\bar{f})\right) = W_{R(t)}(f).
\end{eqnarray*}
This proves that $\e^{\i tH_I}$ is a Bogoliubov dynamics implementing
$R(t).$ And hence $\e^{\i tH_I}$ and $U_I(t)$ are equal up to a phase
factor. In order to prove that this phase is one, we will show that
they have the same natural cocycle. By (\ref{eq:uutildephase}), we know that 
\[
U_I(t)=\e^{-\frac{\i}{2}\Re(\int_0^t \Tr(v\bar{L}(s))\d s)}U_\nat(t).
\]

Let now $\rho(t)\in\rr$ be such that $U_\nat(t)=\e^{\i\rho(t)}\e^{\i tH_I}.$ Note that $\Omega\in\cD,$ hence,
\[
\e^{\i\rho(t)}=\frac{\langle \Omega|U_\nat(t) \Omega\rangle}{\langle
  \Omega| \e^{\i tH_I}\Omega\rangle}
\]
is continuously differentiable by Lemma \ref{prop:u(r)w-diff}. Moreover, for all $t,$ 
$\langle \Omega|U_\nat(t) \Omega\rangle \in \rr,$ thus
\[
\frac{\d}{\d t} \langle \Omega|U_\nat(t) \Omega\rangle =
\i\rho'(t)\langle \Omega|U_\nat(t) \Omega\rangle +\langle U_\nat(t)^*
\Omega|\i H_I \Omega\rangle \ \in \rr,
\]
and hence
\beq\label{eq:last}
\rho'(t)\langle \Omega|U_\nat(t) \Omega\rangle=-\Im \langle
U_\nat(t)^*\Omega|\i H_I\Omega\rangle=-\frac{1}{2}\Im \langle U_\nat(t)^*\Omega|\i a^*(v) \Omega\rangle.
\eeq
Therefore
\[
\rho'(t)=-\Im \frac{\langle U_\nat(t)^*\Omega|\i a^*(v)
  \Omega\rangle}{2\langle \Omega|U_\nat(t) \Omega\rangle}=-\frac{1}{2}\Im \langle
e^{-\frac{1}{2}a^*(L(t))}\Omega| \i a^*(v)\Omega\rangle=
\frac{1}{4}\Im \langle a^*(L(t))\Omega| \i a^*(v)\Omega\rangle. 
\]
Now, using (\ref{ccr3}), we have
\[
\langle \Omega| [a(L(t)),a^*(v)]\Omega\rangle =2\Tr (L(t)^*v).
\]
But $L(t)^*=\bar{L}(t),$ therefore
\beq\label{eq:rho}
\rho(t)=\frac{1}{2} \int_0^t \Re \Tr (v\bar{L}(s)) \d s,
\eeq
and 
\[
\quad\qquad\qquad\qquad\qquad\qquad\qquad\qquad\e^{\i tH_I}=\e^{-\frac{\i}{2} \Re(\int_0^t \Tr (v\bar{L}(s)) \d s)}U_\nat(t).\qquad\qquad\qquad\qquad\qquad\qquad\qquad\Box
\]


\subsection{Infimum of Bogoliubov Hamiltonians}\label{ssec:type2}

In this section, we introduce another (natural) distinguished class of
Bogoliubov Hamiltonians, those whose infimum is zero.

\bed\label{def:sympltype2} A unitarily implementable
symplectic group is of type II if and only if it has a bounded
from below Bogoliubov Hamiltonian (and hence all its Bogoliubov
Hamiltonians are bounded from below).
\eed

\bed\label{def:hamtype2} If $R(t)$ is a symplectic group of type II,
we define the Bogoliubov Hamiltonian of type II to be the unique associated
Bogoliubov Hamiltonian whose infimum of spectrum is $0.$ We denote it
by $H_{II}$. The corresponding Bogoliubov unitary group is denoted
$U_{II}(t)=\e^{\i tH_{II}}.$
\eed

We denote by $\cY^\#$ the dual space of $\cY$.

\bed\label{def:classsymb} The classical symbol associated to a one
parameter symplectic group $R(t)=\exp t\left(\begin{array}{cc} \i h &
    -\i v \\ \i \bar{v} & -\i\bar{h} \end{array}\right)$ is the
bilinear symmetric form defined on $\cY^\#$ as 
$$
\cY^\#\times\cY^\# \ni ((\bar{f},f),(\bar{g},g)) \mapsto \frac{1}{2}(\langle f|hg\rangle+\langle
\bar{f}|\bar{h}\bar{g}\rangle+\langle f|v\bar{g}\rangle+\langle
\bar{f}|\bar{v} g\rangle).
$$
\eed

\bet\label{thm:formulainf} Suppose $\fh$ is finite
dimensional. Then every symplectic group with a positive classical
symbol is both of type I and II. Moreover, the Bogoliubov
Hamiltonians $H_I$ and $H_{II}$ satisfy
\begin{equation}\label{eq:formulainf}
H_{II}=H_I-\frac{1}{4}\Tr \left[ \left( \begin{array}{cc} \bar{h}^2-\bar{v}v & \bar{h}\bar{v}-\bar{v}h
 \\ hv-v\bar{h}  & h^2-v\bar{v}  \end{array}\right)^{1/2}-\left( \begin{array}{cc} \bar{h} & 0 \\ 0
 & h \end{array}\right) \right].
\end{equation}
\eet

\proof The operator $v$ is Hilbert-Schmidt (we are in finite
dimension), hence by Theorem \ref{thm:selfadj2} $R(t)$ is of type I and $H_I=\d\Gamma(h)+\frac{1}{2}(a(v)+a^*(v))$.

Let $d$ denote the (complex) dimension of $\fh$, hence $\cY$ is a real
symplectic space of dimension $2d$. We define, on $\cY^\#$, the operator
$$
\beta(h,v):= \frac{1}{2}\left(\begin{array}{cc}  v & h \\ \bar{h} & \bar{v} \end{array}\right).
$$
The operator $\beta(h,v)$ is real symmetric and hence induces a real
quadratic form on $\cY^\#$
$$
\cY^\#\ni (\bar{f},f)\mapsto \langle (f,\bar{f})|\beta(h,v)(\bar{f},f)\rangle,
$$
which is nothing else but the classical symbol of $R(t).$
Its Weyl quantization, denoted $\Op(\beta),$
is then 
\begin{equation}\label{def:quadquant}
\Op(\beta)=\d\Gamma(h)+\frac{1}{2}a^*(v)+\frac{1}{2}a(v) +\frac{1}{2}\Tr(h).
\end{equation}
We also denote  
$$
\sigma=\left(\begin{array}{cc} 0 & \i \\ -\i & 0  \end{array}\right).
$$ 
The map $\cY\times\cY \ni(y,y')\mapsto \frac{1}{2}\langle
\bar{y}|\sigma y'\rangle=\sigma(y,y')$ is the symplectic form on $\cY$ introduced in Section \ref{ssec:fockrepr}.

Since $\beta$ is positive real symmetric and $\sigma$ is real
antisymmetric, we can diagonalize them simultaneously, i.e. there is a basis $(y_1,\cdots,y_{2d})$ of $\cY$ and positive real numbers
$\lambda_1,\cdots,\lambda_{2d}$ such that 
\begin{equation}\label{eq:beta}
\beta \bar{y}_{j}=\lambda_j y_j,
\end{equation}
\begin{equation}
\sigma y_{2j-1}=\bar{y}_{2j}, \quad \sigma y_{2j}=-\bar{y}_{2j-1},\label{eq:sigma}
\end{equation}
where, if $y=(f,\bar{f})$, $\bar{y}=(\bar{f},f).$
Let $f_k\in\fh$ be such that $y_k=(f_k,\bar{f}_k),$ and let
$h_k=|f_k\rangle\langle f_k|$ and $v_k=|f_k\rangle\langle\bar{f}_k|.$
Finally, we denote $\beta_k=\beta(h_k,v_k).$ One then gets,
using (\ref{eq:beta}), 
$\beta(h,v)=\displaystyle{\sum_{j=1}^{2d}} \lambda_j\beta_j.$
Hence 
$$
\Op(\beta(h,v))=\sum_{j=1}^{2d} \lambda_j\Op(\beta_j),
$$
with 
$$
\Op(\beta_j)=\d\Gamma(|f_j\rangle\langle f_j|)+\frac{1}{2}(a(f_j\otimes
f_j)+a^*(f_j\otimes f_j))+\frac{1}{2}=\phi(f_j)^2,
$$
and where $\phi(f)$ denotes the field operators (Section
\ref{ssec:generfock}), so that
\begin{equation}\label{eq:beta2}
\Op(\beta(h,v))=\sum_{j=1}^d (\lambda_{2j-1}\phi(f_{2j-1})^2+\lambda_{2j}\phi(f_{2j})^2).
\end{equation}

Now, since $(y_1,\cdots,y_{2d})$ diagonalizes $\sigma$, we have
$$
\sigma(y_{2j},y_{2k})=\sigma(y_{2j-1},y_{2k-1})=0, \quad {\rm and} 
\quad \sigma(y_{2j},y_{2k-1})=\delta_{jk}.
$$
And hence, $[\phi(f_{2j-1}),\phi(f_{2j})]=\i$ for all
$j\in\{1,\dots,d\}$ while the other field operators
commute. Therefore, by (\ref{eq:beta2}), and the properties of the
harmonic oscillator, 
$$
\inf \Op(\beta(h,v))=\sum_{j=1}^d \sqrt{\lambda_{2j-1}\lambda_{2j}}.
$$

On the other hand, using (\ref{eq:beta})-(\ref{eq:sigma}), one gets  
$$
-(\sigma\beta)^2\bar{y}_{2j-1}=\lambda_{2j-1}\lambda_{2j}\bar{y}_{2j-1}, \quad -(\sigma\beta)^2\bar{y}_{2j}=\lambda_{2j-1}\lambda_{2j}\bar{y}_{2j}.
$$
Therefore we have  
$$
\inf \Op(\beta(h,v))=\sum_{j=1}^d \sqrt{\lambda_{2j-1}\lambda_{2j}}=\frac{1}{2}\Tr \sqrt{-(\sigma\beta)^2}.
$$
Finally, a simple computation gives $-(\sigma\beta)^2=\frac{1}{4}\left(
  \begin{array}{cc} \bar{h}^2-\bar{v}v & \bar{h}\bar{v}-\bar{v}h \\
    hv-v\bar{h} & h^2-v\bar{v}  \end{array}\right),$
from which (\ref{eq:formulainf}) follows since $H_I=\Op(\beta(h,v))-\frac{1}{2}\Tr(h).$
\hfill\qed


\subsection{Relative boundedness of quadratic annihilation and creation operators}\label{sssec:sarelbound}

In this section, we consider $a^*(v)+a(v)$
as a perturbation of $\d\Gamma(h)$ and derive a condition so that it is relatively
bounded with respect to it. 
\bet\label{thm:relbound} If $h$ is a positive selfadjoint operator on $\fh$ and $v \in
\FDom (h^{-1/2}\otimes h^{-1/2})\cap \FDom (h^{-1/2}\otimes 1+1\otimes
h^{-1/2}),$ then $a(v)+a^*(v)$ is $\d \Gamma (h)$ bounded with
relative bound less than $2\|(h^{-1/2}\otimes h^{-1/2})v\|.$
\eet

Using the Kato-Rellich Theorem (\cite{RS2}, Theorem X.39), one then
immediately gets

\bec\label{coro:selfadj1} Under the same assumption, if moreover
$\|(h^{-1/2}\otimes h^{-1/2})v\|< 1$ then the Bogoliubov Hamiltonian
$H:=\d\Gamma (h)+ \frac{1}{2}(a(v)+a^*(v))$ is selfadjoint on
$\FDom (\d\Gamma(h))$ and bounded from below. In particular, the
associated symplectic group $R(t)$ is both of type I and type II.
\eec

The above condition should not be so surprising. Indeed, it closely resembles the condition one can find for the Van-Hove and the Pauli-Fierz Hamiltonians (see \emph{e.g.} \cite{De,DJ}), where a perturbation linear in the annihilation and creation operators is involved.

\bel\label{lem:equivadef} Suppose that $v \in \FDom (h^{-1/2}\otimes h^{-1/2}).$ Then, there exist orthonormal bases of $\fh$ $(\xi_n)_n,(\chi_n)_n$ and positive numbers $\mu_n$ such that $(h^{-1/2}\otimes h^{-1/2})v= \sum_n \mu_n \ \xi_n\otimes\chi_n.$ Moreover, for all $\Psi \in \Gamma^{{\rm fin}}_\s(\fh),$ 
\[
a(v)\Psi=\sum_n \mu_n a(h^{1/2}\xi_n)a(h^{1/2}\chi_n)\Psi.
\]
\eel

\proof It follows from (\ref{eq:defdoubleaa*}) and the fact that $v=\sum \mu_n \ h^{1/2}\xi_n\otimes h^{1/2}\chi_n.$
\hfill\qed

We now prove bounds on $a(v)$ and $a^*(v)$ which generalise the ones obtained in Proposition \ref{prop:aa*dom}, and which are in the spirit of the $N_\tau-$estimate of Proposition \ref{prop:abound}.

\bep\label{prop:arelbound} Suppose $v\in \FDom (h^{-1/2}\otimes h^{-1/2}).$ Then for all $\Psi\in \FDom (\d\Gamma(h)),$ 
\[
\| a(v)\Psi\| \leq \|(h^{-1/2}\otimes h^{-1/2})v\| \|\d\Gamma (h)\Psi\|.
\]
\eep

\proof Using Lemma \ref{lem:equivadef}, we have
\[
\| a(v) \Psi\|^2 = \| \sum_n \mu_n a(h^{1/2}\xi_n)a(h^{1/2}\chi_n)\Psi\|^2 \leq \sum_n \mu_n^2 \, \, \sum_n \| a(h^{1/2}\xi_n)a(h^{1/2}\chi_n)\Psi\|^2. 
\]
Hence, using $\sum_n \mu_n^2=\|(h^{-1/2}\otimes h^{-1/2})v\|^2$ and
Proposition \ref{prop:abound}, we get 
\begin{eqnarray*}
\|a(v) \Psi\|^2 & \leq & \|(h^{-1/2}\otimes h^{-1/2})v\|^2 \sum_n \langle a(h^{1/2}\chi_n)\Psi| \d\Gamma(h) a(h^{1/2}\chi_n)\Psi\rangle \\
 & = & \|(h^{-1/2}\otimes h^{-1/2})v\|^2 \sum_n \left( \langle a(h^{1/2}\chi_n)\Psi| a(h^{1/2}\chi_n) \d\Gamma(h) \Psi\rangle - \langle a(h^{1/2}\chi_n)\Psi| a(h^{3/2}\chi_n)\Psi\rangle \right)\\
 & = & \|(h^{-1/2}\otimes h^{-1/2})v\|^2 \left( \|\d\Gamma (h)\Psi\|^2 -\langle \Psi| \d\Gamma(h^2) \Psi\rangle \right),
\end{eqnarray*}
where in the last line we used the following identities
\[
\qquad\qquad\qquad\sum a^*(h^{1/2}\chi_n)a(h^{1/2}\chi_n)=\d\Gamma(h), \quad {\rm and} \quad
\sum a^*(h^{1/2}\chi_n)a(h^{3/2}\chi_n)=\d\Gamma(h^2).\qquad\qquad  \Box
\]

\bep\label{prop:a*relbound} Suppose $v\in \FDom (h^{-1/2}\otimes h^{-1/2})\cap \FDom (h^{-1/2}\otimes 1+1\otimes h^{-1/2}).$ For any $\epsilon>0,$ there exists $C_\epsilon>0,$ such that for all $\Psi \in \FDom (\d\Gamma(h)),$
\[
\|a^*(v) \Psi\|^2\leq \left( \|(h^{-1/2}\otimes h^{-1/2})v\|^2 +\epsilon \right)\|\d\Gamma(h)\Psi\|^2 +C_\epsilon \|\Psi\|^2.
\]
\eep

In order to prove this estimate, we will need the following lemma which follows directly from (\ref{ccr3}).

\bel\label{lem:aa*normrel} Let $v \in \Gamma_\s^2(\fh),$ then for all $\Psi\in \FDom(N),$
\[
\|a^*(v)\Psi\|^2=\|a(v)\Psi\|^2+4\langle \Psi| \d\Gamma(v v^*) \Psi\rangle+2\|v\|^2\|\Psi\|^2.
\]
\eel

\noindent {\bf Proof of Proposition \ref{prop:a*relbound}}\ \ Using Proposition \ref{prop:arelbound} and Lemma \ref{lem:aa*normrel}, it suffices to show that 
\[
\langle \Psi, \d\Gamma (v v^*)\Psi\rangle \leq \epsilon \|\d\Gamma(h)\Psi\|^2 +C_\epsilon' \|\Psi\|^2
\]
for some $C_\epsilon'.$
One can write $v v^*= h^{1/2} (h^{-1/2}v)(h^{-1/2}v)^* h^{1/2}.$ Now,
$h^{-1/2}v$ is bounded. It is actually Hilbert-Schmidt since $v\in
\FDom (h^{-1/2}\otimes 1+1\otimes h^{-1/2})$. Thus
$v v^*\leq \|h^{-1/2}v\|^2\, h,$ and so 
\[
\langle \Psi | \d\Gamma(v v^*)\Psi\rangle \leq \|h^{-1/2}v\|^2 \|\d\Gamma(h)^{1/2}\Psi\|^2,
\]
which ends the proof. 
\hfill\qed

\noindent {\bf Proof of Theorem \ref{thm:relbound}}\ \ It follows
directly from Propositions  \ref{prop:arelbound} and \ref{prop:a*relbound}.
\hfill\qed


\section{A concrete example: the diagonal case}\label{sec:example}

The case where the one particle space $\fh$ is finite dimensional is
completely understood: all symplectic groups are of type I and we have
a necessary and sufficient condition on its generator to determine
wether it is of type II or not (Section \ref{ssec:type2}).
In this section we consider the simplest ``infinite dimensional''
case. Namely, $\fh:= L^2(\nn)$ with its canonical basis
$(e_n)_{n\in\nn}$ and $h$ and $v$ are both diagonal, i.e.
\beq\label{def:hvdiag}
h:= \sum_n h_n |e_n\rangle\langle e_n|, \qquad v:= \sum_n v_n |e_n\rangle\langle e_n|,
\eeq
and where the $h_n$ are real numbers so that $h$ is selfadjoint.
Our goal is to describe, in this simple situation, what are the one parameter
symplectic groups $R(t)$ which are unitarily implementable, which are
those of type I, and those of type II.
In the case where $R(t)$ is not of type I, we
will also achieve the ``phase renormalization'' we have mentioned in
the introduction. More precisely, we
will prove the following 
\bet\label{thm:example} Consider on $L^2(\nn)$ the operators $h$ and
$v$ defined by (\ref{def:hvdiag}). 
\begin{enumerate} 
\item[(i)] $R(t)$ defines a strongly continuous one parameter group of
      symplectic maps if and only if $v$ is $h$-bounded with relative
      bound strictly less than one, i.e. there exists $a\in[0,1[$ and
      $b\geq 0$ such that for all $n\in\nn,$ $|v_n|\leq a|h_n|+b$.
\item[(ii)] $R(t)$ is unitarily implementable if and only if $\sum
      \frac{|v_n|^2}{1+h_n^2}<+\infty.$ If it is unitarily
      implementable, the
      operators
$$
U_\ren(t):=\e^{\i\Tr(\frac{1}{2}\Re\int_0^t Q_\tau
  v\bar{P}_\tau^{-1}\d\tau+\Lambda_{{\rm ren}}t)}U_\nat(t),
$$
where $\Lambda_\ren=\sum_{|h_n|>1} \frac{|v_n|^2}{4h_n}
|e_n\rangle\langle e_n|,$ form a Bogoliubov dynamics implementing $R(t).$
\item[(iii)] A unitarily implementable symplectic group $R(t)$ is of type I if and only if $\sum \frac{|v_n|^2}{1+|h_n|}<+\infty.$
\item[(iv)] A unitarily implementable symplectic group $R(t)$ is of
      type II if and only if $h_n\geq |v_n|$ for all $n$ and
      $\sum_{|h_n|\leq 1} \frac{|v_n|^2}{|h_n|}<+\infty.$
\end{enumerate}
\eet

Suppose now that $(O_1,\cdots,O_M)$ is a partition of $\nn$, then we
have $\fh= \fh_1 \oplus \cdots \oplus \fh_M$ where $\fh_j={\rm Span}
\{e_n,n\in O_j\}$. But, since $h$ and
$v$ are both diagonal they leave the $\fh_j$ invariant, and we can split the
problem with respect to the above decomposition of $\fh$,
i.e. the symplectic group $R(t)$ can be written as $R(t)=R_1(t)\oplus
\cdots \oplus R_M(t)$ where for all $j=1,\cdots,M$ $R_j(t)$
is a symplectic group on $\cY_j=\{(f,\bar{f}),f\in\fh_j\},$ and we can
consider separately the $M$ so obtained reduced problems. It is then easy to see
that $R(t)$ is unitarily implementable if and only the $R_j(t)$ are
all unitarily implementable, and that the same statement holds for the
type I (resp. type II) character of $R(t)$. For that reason, as a
first step we will consider the case of a single degree of freedom,
i.e. the case $\fh=\cc$.


\subsection{Bogoliubov transformations of a single degree of freedom}\label{ssec:onedim}

Let $\fh=\cc$ and $A=\i\left(\begin{array}{cc} h & -v \\ \bar{v} &
    -\bar{h} \end{array}\right),$ where $h\in\rr$ and $v\in\cc$. 
One can compute explicitly the operators $P(t)$
and $Q(t):$ 
\begin{itemize}
\item if $|h|<|v|$
\beq\label{formulapq1}
P(t)=\cosh(t\sqrt{|v|^2-h^2})+\i
h\frac{\sinh(t\sqrt{|v|^2-h^2})}{\sqrt{|v|^2-h^2}} \quad {\rm
  and}\quad Q(t)=\i
\bar{v}\frac{\sinh(t\sqrt{|v|^2-h^2})}{\sqrt{|v|^2-h^2}},
\eeq
\item if $|h|=|v|$
\beq\label{formulapq2}
 P(t)=1+\i th \quad {\rm and} \quad Q(t)=\i t\bar{v},
\eeq
\item if $|h|>|v|$
\beq\label{formulapq3}
P(t)=\cos(t\sqrt{h^2-|v|^2})+\i
h\frac{\sin(t\sqrt{h^2-|v|^2})}{\sqrt{h^2-|v|^2}} \quad {\rm and} \quad Q(t)=\i
\bar{v}\frac{\sin(t\sqrt{h^2-|v|^2})}{\sqrt{h^2-|v|^2}}.
\eeq
\end{itemize}
From Theorem \ref{thm:formulainf} we know that $R(t)$ is
always of type I with 
$H_I=\d\Gamma(h)+\frac{1}{2}(a^*(v)+a(v)).$
Moreover, it is of type II if and only if its classical
symbol is positive i.e. $\forall z\in\cc, \Re (h|z|^2+vz^2)\geq 0$,
which is equivalent to $h\geq |v|$. The Bogoliubov Hamiltonian of type
II then writes, according to (\ref{eq:formulainf}), 
\beq\label{1diminf}
H_{II}=H_I-\frac{1}{2}(\sqrt{h^2-|v|^2}-h).
\eeq


\subsection{Proof of Theorem \ref{thm:example}}\label{ssec:proofex}

We now turn back to the general situation (\ref{def:hvdiag}).  
In view of (\ref{formulapq1})-(\ref{formulapq2})-(\ref{formulapq3}),
we consider the following partition of 
$\nn:=\cN_< \cup \cN_=\cup \cN_>,$
where $\cN_<:=\{n\in\nn, |h_n|<|v_n| \}$, $\cN_= :=\{n\in\nn, |h_n|=|v_n| \}$ and
$\cN_>:=\{n\in\nn, |h_n|>|v_n|\}$ and split the analysis with
respect to this partition. It will also be convenient to
split again the case $|h|>|v|$ in two cases, namely
$|h_n|^2-|v_n|^2\leq \frac{1}{2}$ and $|h_n|^2-|v_n|^2>\frac{1}{2}$ (the
choice of the value $\frac{1}{2}$ is purely arbitrary and could be
replaced by any strictly positive number).


\subsubsection{The case $|h|<|v|$}\label{ssec:h<v}

Throughout this section we assume that, for all $n$, $|h_n|<|v_n|.$
\bep\label{prop:h<v}  
\begin{itemize}
\item[(i)] $R(t)$ defines a strongly continuous symplectic group if and only
if $v$ is bounded.
\item[(ii)] $R(t)$ is unitarily implementable if and only if $v$ is Hilbert-Schmidt.
\item[(iii)] All unitary implementable symplectic groups are of type I.
\item[(iv)] A unitarily implementable symplectic group is never of
      type II.
\end{itemize}
\eep
\proof $(i)$ If the operator $v$ is bounded then the result follows from
Proposition \ref{prop:gensympl}. Suppose now that $R(t)$ defines a
strongly continuous group. Then there exist two constants $M$ and
$\omega$ strictly positive such that, for all $t,$ $\|R(t)\|\leq
M\e^{\omega |t|}.$ Then, using (\ref{formulapq1}), one easily gets that the operators
$\sqrt{|v|^2-h^2}$ and $\frac{v}{\sqrt{|v|^2-h^2}}$ have to be bounded
which implies that $v$ is bounded.

$(ii)$ Once again the sufficient condition follows from the general
theory (Theorem \ref{thm:genshale}). Suppose now that $R(t)$ is
unitarily implementable. Then, by Theorem \ref{thm:abstunitimpl}, $Q(t)$ is Hilbert-Schmidt for all $t$, i.e. 
$$
\sum \left||v_n|^2\frac{\sinh^2(t\sqrt{|v_n|^2-h_n^2})}{|v_n|^2-h_n^2}\right|^2<\infty.
$$
The result follows immediately since for all $x,$ $\sinh^2(x)\geq x^2.$ 

$(iii)$ This follows from Theorem \ref{thm:gentype1} $(ii)$.

$(iv)$ Since $|h_n|<|v_n|$, the result follows from the properties of ``one degree of freedom'' case. 
\hfill\qed


\subsubsection{The case $|h|=|v|$}\label{ssec:h=v}

We suppose in this section that for all $n$, $|h_n|=|v_n|.$ This situation is very
close to the previous one. 
\bep\label{prop:h=v} 
\begin{itemize}
\item[(i)] $R(t)$ defines a strongly continuous symplectic group if and only
if $v$ is bounded.
\item[(ii)] $R(t)$ is unitarily implementable if and only if $v$ is Hilbert-Schmidt. 
\item[(iii)] All unitarily implementable symplectic groups are of type
      I.
\item[(iv)] A unitarily implementable symplectic group is of type II if and only if $h\geq 0$ and is trace
      class. If it is of type II then $H_{II}=H_I+\frac{\Tr (h)}{2}.$
\end{itemize}
\eep

\proof The proofs of $(i)$-$(ii)$-$(iii)$ are the same as in the previous
section. It remains to prove $(iv)$. Since $v$
is Hilbert-Schmidt, we know
that $R(t)$ is of type I with Bogoliubov Hamiltonian $H_I$ given by
(\ref{intro:ham}) with $c=0$.
Thus $R(t)$ is of type II if and only if $H_I$ is bounded from below.

First assume that $h$ is positive and trace class. Formally, $H_I$ is given by
$H_I=\sum_n H_n,$
where $H_n=\d\Gamma(h_n|e_n\rangle\langle
e_n|)+\frac{1}{2}a^*(v_n|e_n\rangle\langle
e_n|)+\frac{1}{2}a(v_n|e_n\rangle\langle
e_n|).$ Note that the $H_n$ commute with one
another. We shall prove that $H(N):=\sum_{n\leq N} H_n$ converges to
$H_I$ in the strong resolvent sense when $N$ goes to infinity. If this
holds, we have then (\cite{RS1}, Theorem VIII.24) 
$$
\inf H_I \geq \lim_{N\to\infty} \inf H(N). 
$$
On the other hand, $H(N)$ is bounded from below for all $N$ (Section
\ref{ssec:onedim}) and 
\beq\label{eq:HNinf}
\inf H(N)=-\frac{1}{2}\sum_{n=0}^N h_n-(h_n^2-|v_n|^2)^{1/2}=-\frac{1}{2}\sum_{n=0}^N h_n.
\eeq 
Since $h$ is positive and trace class this proves that $H_I$ is bounded from below and 
\beq\label{h=v-inf}
\inf H_I \geq -\frac{\Tr (h)}{2}.
\eeq 

Since $H(N)$ is selfadjoint for all $N$ (Section \ref{ssec:quadham}), 
to prove that $H(N)$ converges to $H_I$ in the strong resolvent
sense it suffices to prove the strong convergence of the unitary
groups, i.e. $U_N(t):=e^{\i tH(N)}$ converges strongly to $U_I(t)$ for all
$t,$ which is equivalent to prove that $\tilde{U}_N(t):=U_N(t)^{-1}U_I(t)$ strongly
converges to the identity. Moreover it is clearly sufficient to prove
strong convergence on the dense set $\Gamma_\s^\fin(C_c(\nn))$ where
$C_c(\nn)$ denotes the set of sequences which have compact support
(since $C_c(\nn)$ is not a Hilbert space, $\Gamma_\s^\fin(C_c(\nn))$
denotes here, with an abuse of notation, the algebraic Fock space over $C_c(\nn)$). 

Let 
$$
h(N):=\sum_{n>N} h_n|e_n\rangle\langle e_n| \quad {\rm and}\quad v(N):=\sum_{n>N} v_n|e_n\rangle\langle e_n|.
$$
We also denote by $R(t,N)$ the corresponding symplectic group and
similarly for $P(t,N)\dots$
One then easily gets
$$
\tilde{U}_N(t)=\e^{-\frac{\i}{2}\Tr\left(\int_0^t
    Q(s,N)v(N)\bar{P}(s,N)^{-1}
  \right)}\e^{-\frac{1}{2}a^*(K(t,N))}\Gamma((P(t,N)^{-1})^*)\e^{-\frac{1}{2}a(L(t,N))}.
$$
Since  $\int_0^t Q(s)v\bar{P}(s)^{-1}\d s$ is
trace class by Theorem \ref{thm:gentype1}, we have 
\begin{equation}\label{eq:unlim1}
\lim_{N\to +\infty} \e^{\frac{1}{2}\Tr\left(\int_0^t
Q(s,N)v(N)\bar{P}^{-1}(s,N)\d s \right)}=1.
\end{equation}
Moreover, let $\Phi\in\Gamma_\s^\fin(C_c(\nn)),$ then for $N$ large enough one has
\begin{equation}\label{eq:unlim2}
\Gamma((P(t,N)^{-1})^*)\e^{-\frac{1}{2}a(L(t,N))}\Phi=\Phi.
\end{equation}
Finally, since $K(t)$ is Hilbert-Schmidt, the sequence of operators
$K(t,N)$ goes to zero in the Hilbert-Schmidt norm. This together with
Proposition \ref{prop:expa*cvg} proves that
\begin{equation}\label{eq:unlim3}
\lim_{N\to+\infty} \e^{-\frac{1}{2}a^*(K(t,N))}\Phi=\Phi.
\end{equation}
The strong convergence of $\tilde{U}_N(t)$ to the identity on
$\Gamma_\s^\fin(C_c(\nn))$ follows from (\ref{eq:unlim1})-(\ref{eq:unlim2})-(\ref{eq:unlim3}).

We now suppose that $H_I$ is bounded from below. Let
$\fh_N:={\rm Span}\{e_n, n=0,\cdots N\}$. $\Gamma_\s(\fh)$ is isomorphic to $\Gamma_\s(\fh_N) \otimes
\Gamma_\s(\fh_N^\perp)$ and via this identification, and with a
slight abuse of notation, we have
\beq\label{eq:H1decomp}
H(N)=H(N)\otimes 1 \quad {\rm and} \quad 
H_I=H(N)\otimes 1+1\otimes (H_I-H(N))
\eeq
where $H_I-H(N)$ acts on $\Gamma_\s(\fh_N^\perp)$ and is defined as $H_I$ but with
$h\lceil_{\fh_N^\perp}$ and $v\lceil_{\fh_N^\perp}$ instead of $h$ and
$v$.

The positivity of the classical symbol of $H(N)$ then writes 
$$
\forall (z_0,\cdots,z_N)\in \cc^{N+1}, \quad \sum_{n=0}^N
\Re(h_n|z_n|^2+v_nz_n^2)\geq 0.
$$
In particular this implies that the $h_n$ are positive. It remains to
prove that $h$ is trace class. 

Let $\epsilon>0$, there exists $\Psi_N\in\cD(H(N))\subset
\Gamma_\s(\fh_N)$ such that $\langle \Psi_N,H(N)\Psi_N\rangle \leq
\inf H(N)+\epsilon=
-\frac{1}{2}\sum_{n=0}^N h_n+\epsilon.$ Let now $\Phi_N:=\Psi_N\otimes
\Omega_N^\perp$ where $\Omega_N^\perp$ denotes the vacuum of
$\Gamma_\s(\fh_N^\perp).$ Using (\ref{eq:H1decomp}), it is then easy
to see that $\Phi_N\in\cD(H_I)$ and
$$
\langle \Phi_N,H_I \Phi_N\rangle \leq -\frac{1}{2}\sum_{n=0}^N h_n+\epsilon.
$$
Since the above inequality holds for all $N$ and $\epsilon>0$, and
since $H_I$ is bounded from below, this proves that $h$ is trace class
and that 
\beq\label{h=v-inf2}
\inf H_I \leq -\frac{\Tr (h)}{2}.
\eeq
Finally, (\ref{h=v-inf}) and (\ref{h=v-inf2}) prove that $H_{II}=H_I+\frac{\Tr(h)}{2}$.
\hfill\qed

Note that if $h$ is positive but is not trace class, we have an
example of a unitarily implementable group $R(t)$ which has a positive
classical symbol but which is not type II.


\subsubsection{The case $0<|h|^2-|v|^2\leq \frac{1}{2}$}\label{sssec:h-v<1}

In this section we now assume that for all $n$, $0<|h_n|^2-|v_n|^2\leq
\frac{1}{2}$.

\bep\label{prop:h-v<1} 
\begin{itemize}
\item[(i)] $R(t)$ defines a strongly continuous group if and only if
      $v$ is bounded.
\item[(ii)] $R(t)$ is unitarily implementable if and only if $v$ is Hilbert-Schmidt.
\item[(iii)] All unitarily implementable symplectic groups are of type
      I.
\item[(iv)] A unitarily implementable symplectic group is of type II
      if and only if $h\geq 0$ and $|v|^2h^{-1}$ is trace class.
\end{itemize}
\eep

\proof $(i)$ If $v$ is bounded the result follows once again from Proposition \ref{prop:gensympl}.

Suppose now that $R(t)$ is a strongly continuous group. A densely defined closed operator $A$
is the generator of strongly continuous group if and only if \cite{Da} there exists
$M\geq 1$ and $\omega\geq 0$ such that 
\begin{itemize} 
\item $]-\infty,-\omega[\, \cup\, ]\omega,+\infty[\subset \rho(A)$, where
      $\rho(A)$ denotes the resolvent set of $A$, 
\item For all $\lambda\in ]-\infty,-\omega[\cup]\omega,+\infty[$ and
      all $m\in\nn,$ $\|(A-\lambda)^{-m}\|\leq \frac{M}{(|\lambda|-\omega)^m}$.
\end{itemize}
It is easy to see that the operator $A=\i\left(
  \begin{array}{cc} h & -v \\ \bar{v} & -\bar{h} \end{array}
\right)$ is closed and densely defined.
Let us denote $A_n:=\i\left(
  \begin{array}{cc} h_n & -v_n \\ \bar{v_n} & -\bar{h_n} \end{array}
\right)\in M_2(\cc)$.
Since $|h_n|>|v_n|$, $A_n-\lambda$ is invertible for any
$\lambda\in\rr,$ and $\|(A_n-\lambda)^{-1}\|= \frac{\sqrt{\lambda^2+|h_n|^2+|v_n|^2}}{\lambda^2+|h_n|^2-|v_n|^2}.$
A necessary condition so that $A$ generates a strongly
continuous group is thus
\beq\label{resolvbound}
\sup_{n\in\nn} \frac{\sqrt{\lambda^2+|h_n|^2+|v_n|^2}}{\lambda^2+|h_n|^2-|v_n|^2}\leq \frac{M}{|\lambda|-\omega},
\eeq
for some $M\geq 1$, $\omega\geq 0$ and for any $|\lambda|>\omega.$
The boundedness of $v$ follows directly from (\ref{resolvbound}) and the
fact that $|h_n|^2-|v_n|^2\leq \frac{1}{2}$. 

$(ii)$ The sufficient condition follows once again from the general
theory (Theorem \ref{thm:genshale}). Suppose now that $R(t)$ is
unitarily implementable. In particular $Q(t)$ has to be
Hilbert-Schmidt for all $t$, i.e. $\forall t\in\rr$,
\beq\label{unitimp:h-v<1}
\sum \left||v_n|^2\frac{\sin^2(t\sqrt{h_n^2-|v_n|^2})}{h_n^2-|v_n|^2}\right|<+\infty.
\eeq
Take $t=\pi$. Since $0<h_n^2-|v_n|^2\leq \frac{1}{2},$ one has, for all
$n,$ $\sin^2(\pi\sqrt{h_n^2-|v_n|^2})\geq 2(h_n^2-|v_n|^2).$
Inserting this in (\ref{unitimp:h-v<1}) proves that $v$ is Hilbert-Schmidt.

$(iii)$ Once again the result follows from Theorem \ref{thm:gentype1}.

$(iv)$ The proof is the same as for the case $|h_n|=|v_n|$ and using
the fact that $\sum (h_n-\sqrt{h_n^2-|v_n|^2})<+\infty \Leftrightarrow
\sum \frac{|v_n|^2}{|h_n|}<+\infty$.
\hfill\qed


\subsubsection{The case $|h|^2-|v|^2> \frac{1}{2}$}\label{sssec:h-v>1}

Finally, in this section we assume that for all $n$, $|h_n|^2-|v_n|^2>\frac{1}{2}$.

\bep\label{prop:h-v>1} 
\begin{itemize}
\item[(i)] $R(t)$ defines a strongly continuous group if and only if
      $\frac{|v|}{\sqrt{|h|^2-|v|^2}}$ is bounded.
\item[(ii)] $R(t)$ is unitarily implementable if and only if $\frac{|v|}{\sqrt{|h|^2-|v|^2}}$ is Hilbert-Schmidt.
\item[(iii)] A unitarily implementable symplectic group is of type
      I if and only if $|v|^2h^{-1}$ is trace class.
\item[(iv)] A unitarily implementable symplectic group is of type II
      if and only if $h\geq 0$.
\end{itemize}
\eep

\proof $(i)$ Suppose $\frac{|v|}{\sqrt{|h|^2-|v|^2}}$ is bounded. Thus $v$ is $h$-bounded with relative bound strictly less than
one. Writing, as in Section \ref{ssec:clasdyn}, $A=A_0+V$ we get that $V$ is $A_0$
bounded with relative bound strictly less than one. Since $A_0$
generates a strongly continuous group ($A_0$ is antiselfadjoint) this
proves that $A$ generates a strongly continuous group \cite{Da}.

Suppose now that $A$ generates a strongly continuous group. Using the
same argument as in the previous section (see (\ref{resolvbound})), there are
constants $M\geq 1$ and $\omega\geq 0$ such that for all $\lambda>\omega,$ and all $n$,
$$
\frac{1}{\sqrt{2}(\sqrt{\lambda^2+h_n^2}-|v_n|)} 
\leq \frac{\sqrt{\lambda^2+|h_n|^2+|v_n|^2}}{\lambda^2+|h_n|^2-|v_n|^2}\leq \frac{M}{\lambda-\omega},
$$
which one can rewrite as
$$
\lambda^2(2M^2-1)-2\lambda(\sqrt{2}M|v_n|-\omega)+2M^2h_n^2-(\sqrt{2}M|v_n|-\omega)^2\geq
0, \qquad \forall \lambda\geq \omega.
$$
The result follows easily from the above
inequality and the assumption $|h_n|^2-|v_n|^2>\frac{1}{2}$.

$(ii)$ Suppose $\frac{|v|}{\sqrt{|h|^2-|v|^2}}$ is
Hilbert-Schmidt. Therefore so is $vh^{-1},$ and hence,
using the fact that $v$ and $h$ commute, Assumption
\ref{assu:genshalemin} is satisfied, so that $R(t)$ is unitarily implementable by
Theorem \ref{thm:genshale}.

Suppose now that $R(t)$ is unitarily implementable. Hence
the map
$
t\mapsto \Tr(\log(1-K(t)^*K(t)))
$
is continuous (see the proof of Theorem \ref{thm:abstunitimpl}) and thus locally integrable. Using (\ref{formulapq3})
we get
$$
\int_0^T \Tr(\log(1-K(t)^*K(t)))\d t=- \int_0^T \sum_n
\log \left(1+\frac{|v_n|^2}{h_n^2-|v_n|^2}\sin^2(t\sqrt{h_n^2-|v_n|^2})
\right)\d t,
$$

Then, using $(i)$, we know that the sequence
$\frac{|v_n|^2}{h_n^2-|v_n|^2}$ is bounded. Hence there
exists $C>0$ such that for all $n\in\nn$ and $t\in\rr$,
$$
\log \left(1+\frac{|v_n|^2}{h_n^2-|v_n|^2}\sin^2(t\sqrt{h_n^2-|v_n|^2})
\right) \geq C \frac{|v_n|^2}{h_n^2-|v_n|^2}\sin^2(t\sqrt{h_n^2-|v_n|^2}),
$$
and hence 
$$
\sum \int_0^T \frac{|v_n|^2}{h_n^2-|v_n|^2}\sin^2(t\sqrt{h_n^2-|v_n|^2}) \d t = \sum \frac{|v_n|^2}{h_n^2-|v_n|^2} \left(
   \frac{T}{2}-\frac{\sin(2T\sqrt{h_n^2-|v_n|^2})}{4\sqrt{h_n^2-|v_n|^2}}\right) <+\infty,
 \quad {\rm for \, all}\, T.
$$
Using $\sqrt{h_n^2-|v_n|^2}>\frac{1}{2}$ and choosing $T$ large enough, we get $\sum \frac{|v_n|^2}{h_n^2-|v_n|^2}<+\infty.$

$(iii)$ If $|v|^2h^{-1}$ is trace class, then Assumption
\ref{assu:gengroup} is satisfied so $R(t)$ is of type I.
Suppose now that $R(t)$ is of type I. Then by definition, $P(t)\e^{-\i
  th}-1$ is trace class for all $t$. Using (\ref{eq:rintegral}) and the fact that
all the operators involved here commute one gets
$P(t)\e^{-\i th}=\e^{\i\int_0^t \bar{Q}(s)\bar{v}P(s)^{-1}\d s}.$
Therefore we have, for all $t$, 
\beq
\sum_n \left|\int_0^t \bar{Q}_n(s)\bar{v}_nP_n(s)^{-1}\d s\right|<+\infty.
\eeq 
Using (\ref{formulapq3}) we get 
\begin{eqnarray}\label{ex:phase1}
\bar{Q}_n(s)\bar{v}_nP_n(s)^{-1} & = &
-h_n|v_n|^2\frac{\sin^2(s\sqrt{h_n^2-|v_n|^2})}{h_n^2-|v_n|^2\cos^2(s\sqrt{h_n^2-|v_n|^2})}\\
 & & \qquad - \i |v_n|^2\sqrt{h_n^2-|v_n|^2}\frac{\sin(s\sqrt{h_n^2-|v_n|^2})\cos(s\sqrt{h_n^2-|v_n|^2})}{h_n^2-|v_n|^2\cos^2(s\sqrt{h_n^2-|v_n|^2})},\nonumber
\end{eqnarray}
so that, in particular, 
\beq
\sum \left| h_n|v_n|^2 \int_0^t \frac{\sin^2(s\sqrt{h_n^2-|v_n|^2})}{h_n^2-|v_n|^2\cos^2(s\sqrt{h_n^2-|v_n|^2})}\d s  \right|<+\infty.
\eeq
The above integral can be explicitly computed and one
gets
\begin{eqnarray}\label{ex:phase3}
 & & h_n|v_n|^2\int_0^t\frac{\sin^2(s\sqrt{h_n^2-|v_n|^2})}{h_n^2-|v_n|^2\cos^2(s\sqrt{h_n^2-|v_n|^2})}\d s\\
 & = & th_n\left(1-\sqrt{1-\frac{|v_n^2|}{h_n^2}}\right)
 +\frac{h_n}{|h_n|}\left[r_n(t)\sqrt{h_n^2-|v_n|^2}-{\rm arccotan} \left(\sqrt{1-\frac{|v_n^2|}{h_n^2}}{\rm cotan} (r_n(t)\sqrt{h_n^2-|v_n|^2}) \right)\right],\nonumber
\end{eqnarray}
where $r=t-\frac{\pi}{\sqrt{h_n^2-|v_n|^2}}{\rm
  E}(\frac{t\sqrt{h_n^2-|v_n|^2}}{\pi})\in
[0,\frac{\pi}{\sqrt{h_n^2-|v_n|^2}}[$ and where ${\rm E}$ denotes the
entire part.

To prove $(iii)$ it suffices to prove that for all $t\in\rr,$
\begin{equation}\label{eq:reste2}
\sum \left|r_n(t)\sqrt{h_n^2-|v_n|^2}-{\rm arccotan} \left(\sqrt{1-\frac{|v_n^2|}{h_n^2}}{\rm cotan} (r_n(t)\sqrt{h_n^2-|v_n|^2}) \right)\right|<+\infty.
\end{equation}
Indeed, if (\ref{eq:reste2}) holds, then one has $\sum \left|
  h_n\left(1-\sqrt{1-\frac{|v_n^2|}{h_n^2}}\right)\right| <+\infty$,
from which the result follows directly using $|h_n|^2>|v_n|^2+\frac{1}{2}\geq \frac{1}{2}$.

Now, it is not difficult to show that for all
$r\in[0,\frac{\pi}{\sqrt{h_n^2-|v_n|^2}}[$ one has
$$
\left|r\sqrt{h_n^2-|v_n|^2}-{\rm arccotan} \left(\sqrt{1-\frac{|v_n^2|}{h_n^2}}{\rm cotan}
      (r\sqrt{h_n^2-|v_n|^2}) \right)\right| 
 \leq
 \frac{\pi}{2}-2\arctan\left(\left(1-\frac{|v_n|^2}{h_n^2}\right)^{1/4}\right).
$$
Since $R(t)$ is unitarily implementable, $\sum
\frac{|v_n|^2}{h_n^2-|v_n|^2}<+\infty$ and hence 
 $\sum \frac{|v_n|^2}{h_n^2}<+\infty$.   
Equation (\ref{eq:reste2}) follows from this and the above majoration.

$(iv)$ We use the notation introduced in the proof of Proposition
\ref{prop:h=v}. Using (\ref{ex:phase1})-(\ref{ex:phase3})-(\ref{eq:reste2}),
one can see that for all $t$
\beq
U_{{\rm ren}}(t):= \e^{\frac{\i}{2} \sum_n\left(\Re\int_0^t Q_n(\tau)
  v_n\bar{P}_n(\tau)^{-1}\d\tau+ th_n\left(1-\sqrt{1-\frac{|v_n|^2}{h_n^2}}\right)\right)} U_\nat(t) 
\eeq
is a well defined Bogoliubov implementer of $R(t)$.

In the same way as in the proof of Theorem \ref{thm:utildegroup}, to
prove that it forms a one paremeter unitary group, it suffices to prove that 
\beq\label{eq:urenphase}
\langle \Omega|U_{{\rm ren}}(t)U_{{\rm ren}}(s)\Omega\rangle=\langle \Omega|U_{{\rm ren}}(t+s)\Omega\rangle.
\eeq
 
Since the operators $h$ and $v$ are both diagonal with respect to the basis $(e_n)_n,$ so are the operators $P(t), Q(t), K(t), L(t).$ Hence one can write both the left and right hand side of (\ref{eq:urenphase}) as a product over $n.$ It therefore suffices to show that, for all $n$,
\begin{eqnarray*}
 & & \exp\left(\frac{1}{2}\int_0^{t_1} Q_n(s) v_n\bar{P}_n(s)^{-1}\d s +
   \i\lambda_nt_1\right)\exp\left(\frac{1}{2}\int_0^{t_2}
   Q_n(s) v_n\bar{P}_n(s)^{-1}\d s + \i\lambda_nt_2\right)\\
 & & \hspace{10cm} \times{\rm det}(1-L^*_n(t_1)K_n(t_2))^{-1/2}\\
 & = & \exp\left(\frac{1}{2}\int_0^{t_1+t_2} Q_n(s)
   v_n\bar{P}_n(s)^{-1}\d s + \i\lambda_n(t_1+t_2)\right),
\end{eqnarray*}
where $\lambda_n=h_n\left(1-\sqrt{1-\frac{h_n^2}{|v_n|^2}}\right).$
This follows from Theorem \ref{thm:gentype1} applied
to the generator $\left( \begin{array}{cc} \i h_n & -\i v_n \\ \i
    \bar{v}_n & -\i h_n \end{array}\right)$ considered on the space $\rr^2.$ 
Hence $R(t)$ is of type II if and only if the generator $H_{{\rm ren}}$ of $U_{{\rm ren}}(t)$ is bounded from
below. 

Suppose $h$ is positive. Let 
$$
H_{II,n}:=H_n-\inf H_n=H_n+\frac{1}{2} h_n\left(1-\sqrt{1-\frac{|v_n|^2}{h_n^2}}\right)
\quad {\rm and} \quad
H_{II}(N):=\sum_{n=0}^N H_{II,n}.
$$
For any $N$, $H_{II}(N)$ is selfadjoint and $\inf
H_{II}(N)=0$.
Moreover, using the same
argument as in the proof of Proposition \ref{prop:h=v}, we prove that
$H_{II}(N)$ converges to $H_{{\rm ren}}$ in the strong resolvent sense
so that
$$
\qquad\qquad\qquad\qquad\qquad\qquad\qquad\qquad \inf H_{{\rm ren}} \geq \lim_{N\to \infty} \inf H_{II}(N)=0. \qquad\qquad\qquad\qquad\qquad\qquad\qquad  \Box
$$


\subsubsection{Putting all together}\label{sssec:endproof}

It is easy to see that in each of the four situations the necessary
and sufficient conditions
which appear in Theorem \ref{thm:example} are equivalent to the
corresponding conditions used in the various propositions. The only
points which are not immediate are the definition of the operator
$\Lambda_\ren$ in $(ii)$ and of the series which appears in $(iv)$.  

The obvious definition of $\Lambda_\ren$ would be to replace the sum over $\{n\;
|\; |h_n|>1\}$  by the same sum but over $\{n\; | \;
|h_n|^2>|v_n|^2+\frac{1}{2}\}$, and similarly for the series in
$(iv)$. To prove that these definitions are equivalent, we have to
prove that $\sum_{n\in\cN} \frac{|v_n|^2}{|h_n|}<+\infty$ where 
$$
\cN=\cN_1\cup\cN_2:=\{ n\;|\; 1< |h_n|^2\leq |v_n|^2+\frac{1}{2} \} \cup \{ n\;|\;
|v_n|^2+\frac{1}{2}<|h_n|^2\leq 1 \}.
$$
Since $R(t)$ is unitarily implementable,we have $\sum_\cN \frac{|v_n|^2}{1+|h_n|^2}<+\infty$.
Using this, it is then clear that $\sum_{n\in\cN_2}
\frac{|v_n|^2}{|h_n|}<+\infty$. On the other hand, if $n\in\cN_1$, one
easily gets that $\frac{|v_n|^2}{1+|h_n|^2}\geq \frac{1}{4}.$ The
implementability of $R(t)$ thus gives that $\cN_1$ is actually a
finite set.
\hfill\qed


\end{document}